\documentclass[conference, compsoc]{IEEEtran}
\IEEEoverridecommandlockouts
\pdfoutput=1

\usepackage[nocompress]{cite}
\usepackage{amsmath,amssymb,amsfonts}
\usepackage{algorithmic}
\usepackage{graphicx}
\usepackage{textcomp}
\usepackage{xcolor}

\usepackage{hyperref}
\usepackage{listings}
\usepackage[labelformat=simple, skip=0pt]{subcaption}

\begin{document}

\title{Exploiting Data Skew for Improved Query Performance
    \textsuperscript{*}\thanks{* This is an extended version of a paper appearing at 36th IEEE International Conference on Data Engineering (ICDE 2020)}
}

\author{
    \IEEEauthorblockN{Wangda Zhang}
    \IEEEauthorblockA{
        \textit{Columbia University}\\
        \textit{zwd@cs.columbia.edu}
    }
    \and
    \IEEEauthorblockN{Kenneth A. Ross\textsuperscript{$\dagger$}\thanks{$\dagger$ Work partly supported by a gift from Oracle Corporation.}}
    \IEEEauthorblockA{
        \textit{Columbia University}\\
        \textit{kar@cs.columbia.edu}
    }
}

\maketitle

\begin{abstract}

Analytic queries enable sophisticated large-scale data analysis within many commercial, scientific and medical domains today. Data skew is a ubiquitous feature of these real-world domains. In a retail database, some products are typically much more popular than others. In a text database, word frequencies follow a Zipf distribution with a small number of very common words, and a long tail of infrequent words. In a geographic database, some regions have much higher populations (and data measurements) than others.

Current systems do not make the most of caches for exploiting skew. In particular, a whole cache line may remain cache resident even though only a small part of the cache line corresponds to a popular data item. In this paper, we propose a novel index structure for repositioning data items to concentrate popular items into the same cache lines. The net result is better spatial locality, and better utilization of limited cache resources. We develop a theoretical model for analyzing the cache behavior, and implement database operators that are efficient in the presence of skew. Our experiments on real and synthetic data show that exploiting skew can significantly improve in-memory query performance. In some cases, our techniques can speed up queries by over an order of magnitude.

\end{abstract}

% \begin{IEEEkeywords}
% component, formatting, style, styling, insert
% \end{IEEEkeywords}

\section{Introduction}
\label{sec:intro}

In online analytic processing (OLAP) a user executes a collection of complex queries over large data sets, in order to understand the data at hand and to obtain actionable knowledge. With the increasing main-memory capacity of contemporary hardware, query execution can occur entirely in RAM. Analytical query workloads that are typically read-only need no disk access after the initial load. In response to this trend, several commercial and research database management systems have been designed (or re-designed) for memory-resident data~\cite{faerber2017main}. Examples of recent systems include H-Store/VoltDB~\cite{kallman2008h}, Hekaton~\cite{diaconu2013hekaton}, HyPer~\cite{kemper2011hyper}, IBM BLINK~\cite{barber2012business}, DB2 BLU~\cite{raman2013db2}, SAP HANA~\cite{farber2012sap}, Vectorwise~\cite{zukowski2012vectorwise}, Oracle TimesTen~\cite{lahiri2013oracle}, MonetDB~\cite{boncz2008breaking}, HYRISE~\cite{grund2010hyrise}, HIQUE~\cite{krikellas2010generating}, LegoBase~\cite{klonatos2014building}, Peloton~\cite{pavlo2017self}, and Quickstep~\cite{patel2018quickstep}. Most analytic database systems use some variant of columnar storage, since only the columns needed to answer the query need to be read~\cite{abadi2013design}.

\begin{figure}
  \begin{subfigure}{0.235\textwidth}
    \centering
    \includegraphics[width=\textwidth]{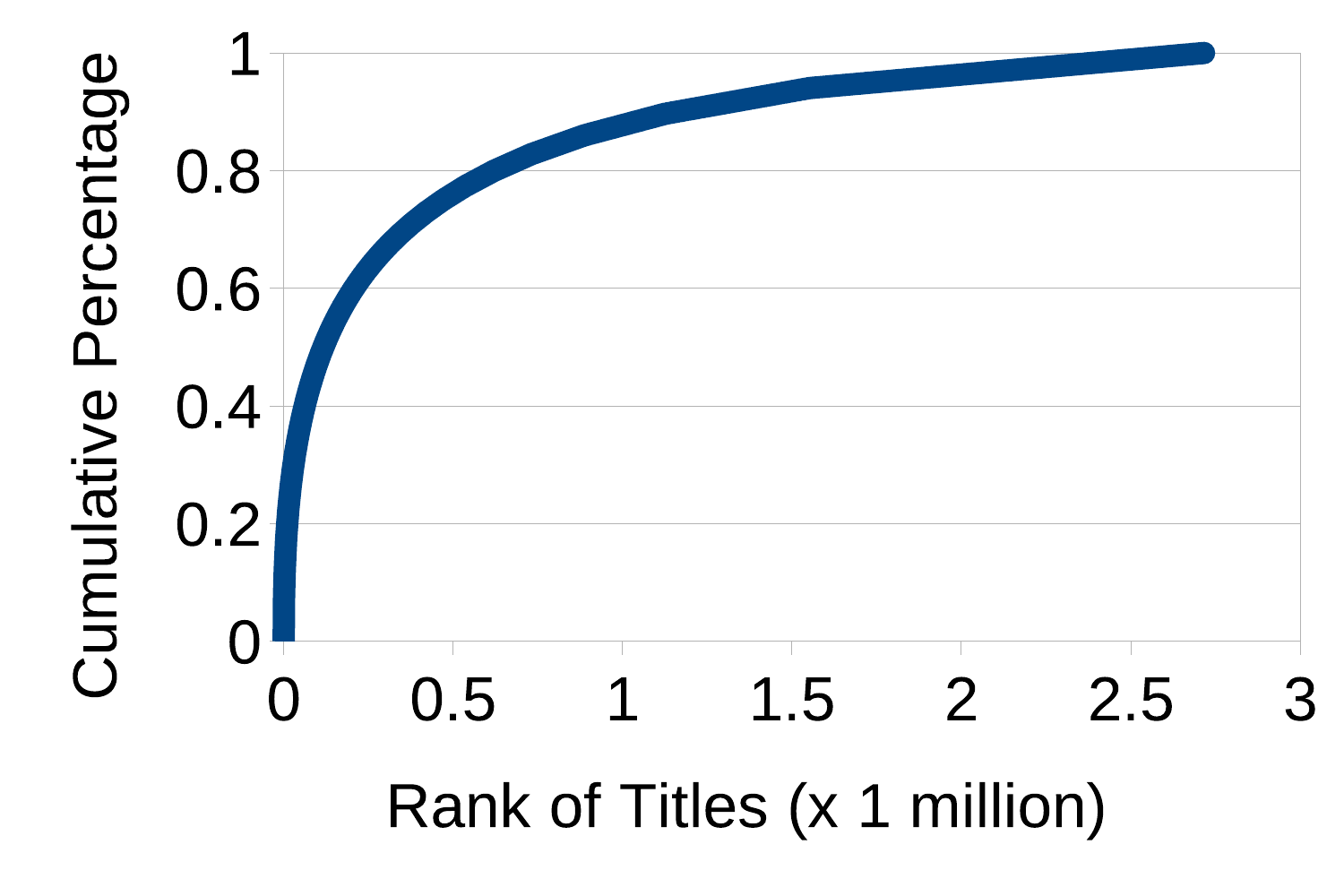}
    \caption{Distribution of book titles}
    \label{fig:distribution:book}
  \end{subfigure}
  \begin{subfigure}{0.235\textwidth}
    \centering
    \includegraphics[width=\textwidth]{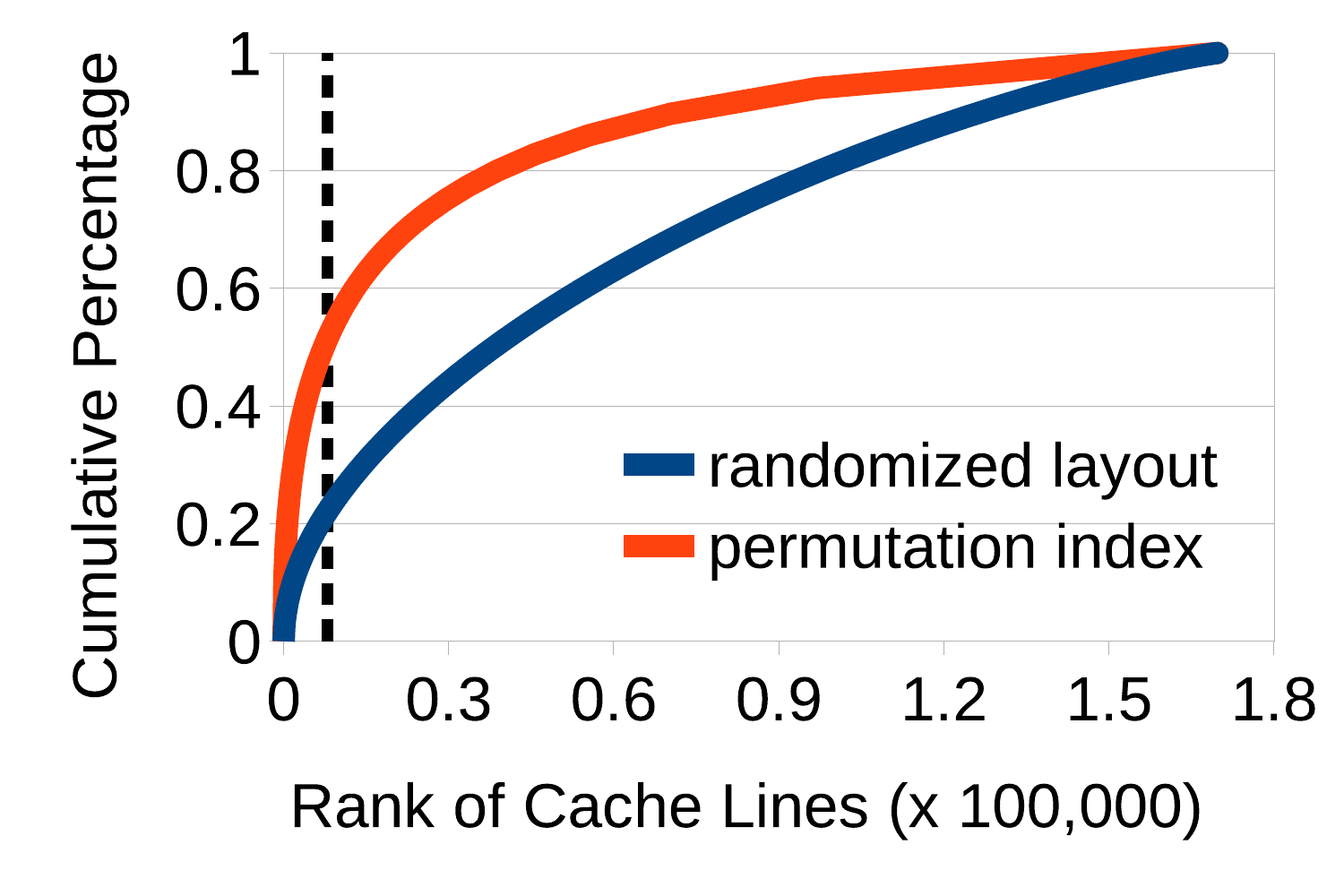}
    \caption{Access to cache lines ({\tt Q1})}
    \label{fig:distribution:cache}
  \end{subfigure}
  \caption{An example of data skew in a real dataset.}
  \label{fig:distribution}
\end{figure}

Skew is a common feature of many real-world domains. Power-law distributions apply to many types of data, including word-usage in text databases, protein interactions, internet routing node-degree, phone-call data, city populations, email contact-lists, surname frequencies, and paper citations~\cite{clauset2009power}. Several other kinds of skewed distributions can also be found~\cite{newman2005power}. Figure~\ref{fig:distribution:book} shows the cumulative distribution of book reviews in the publicly available Amazon reviews dataset, where books are ranked by the number of reviews. The distribution is likely to be a representative proxy of the actual sales data. As shown, the 100,000 most popular book titles ($< 5\%$) account for roughly 50\% of the entire data, and 75\% of the sales concentrate on the top 500,000 titles.

Although skew has been recognized as common and important in empirical database deployments, standard benchmarks such as TPC-H and SSB have specified uniform distributions and avoided skew~\cite{rabl2013variations}. Even when benchmarks such as TPC-DS adopt skew, they are constrained by query comparability issues to make the data more uniform than it would otherwise be~\cite{nambiar2006making}. As a result, there is little research on the impact of skew on analytic query performance.
% and skew-aware data management is not adequately considered in existing main memory database architectures.

In the context of a data warehouse, skew is likely to affect query performance. Consider a large fact table corresponding to the sales of a bookseller. One of the columns of this fact table is the id of a book, represented as an integer foreign key {\tt bid} into a dimension table of detailed book information. Skew as shown in Figure~\ref{fig:distribution:book} in the {\tt bid} column would be expected due to different popularity of the books. The following example queries utilize {\tt bid} in an important way.

\begin{lstlisting}[language=SQL]
-- Q1: Consult a column of the dimension table
Select  B.price
From    Sales S, Books B
Where   S.bid = B.bid
\end{lstlisting}

Query {\tt Q1} executes a foreign key join to read or materialize a column from the dimension table. The obtained prices can then be used to calculate revenues. To answer this query, the database system can scan the fact table, and look up the {\tt price} column of table {\tt B} using {\tt S.bid} as an array offset. If the number of products is such that the {\tt price} column is larger than the cache in size, and {\tt S.bid}s are uniformly distributed in the dimension domain, then this lookup could incur expensive cache misses due to unclustered memory acccesses, which can often be the dominant cost~\cite{schuh2016experimental}. With skew in {\tt S.bid}, the situation is somewhat better because the most frequently occurring items are likely to reside in the cache. Nevertheless, the cache is still underutilized because a single cache-resident cache line will typically hold a small number of popular items and many unpopular items.
% In Section~\ref{sec:idx}, we shall explain this cache behavior in further detail and will propose new ways to concentrate the data so that the cache is better utilized.

% \begin{verbatim}
% // Q1a: Materialize based on the dimension table
% Select  P.price
% From    S, P
% Where   S.pid = P.pid
% \end{verbatim}

% Query {\tt Q1a} materializes a column from the dimension table using foreign keys, as might happen when a schema is denormalized, or when a column is late-materialized from an intermediate result containing identifiers. This materialization requires random reads from the dimension table, and the performance issues are similar to {\tt Q1}.

\begin{lstlisting}[language=SQL]
-- Q2: Filter the fact table based on B.price
Select  S.*
From    Sales S, Books B
Where   S.bid = B.bid and B.price < 100
\end{lstlisting}

To process this selective filtering, the database system can preprocess the dimension table to determine which products meet the {\tt B.price<100} condition. It can then create a bitmap indexed by {\tt bid} that can be consulted using {\tt S.bid} as an offset. {\tt Q2} has similar cache miss issues to {\tt Q1}, except that the data being consulted is one bit per {\tt bid} rather than several bytes. As a result, the product cardinality thresholds for cache-residence will be larger under the same cache capacity.

\begin{lstlisting}[language=SQL]
-- Q3: Aggregate grouped by book ids
Select    bid, count(*)
From      Sales
Group by  bid
\end{lstlisting}

For this query, the database system can create a table of counts, indexed by {\tt bid}, and update the count for each corresponding fact table record in turn. The memory access pattern (and cache behavior) is similar to {\tt Q1}, with the added observation that because cache lines are updated, cache-line replacement triggers some additional memory-write traffic.

\begin{lstlisting}[language=SQL]
-- Q3a: Heavy hitter counts
Select    bid, count(*)
From      Sales
Group by  bid
Order by  count(*) desc
Limit     4000
\end{lstlisting}

Query {\tt Q3a} is similar to {\tt Q3} except that we are only interested in the counts of heavy hitters, which for this query means the {\tt bid}s with a count among the top 4,000 book titles. Because the cache footprint of the heavy hitters is much smaller than the entire {\tt bid} domain, there are opportunities for further performance enhancement if the candidate heavy hitters can be identified in advance.

For these core database operations, skew-aware data management is not adequately considered in existing main memory database architectures. In this paper, we propose a novel index structure called a permutation index for reordering data items by their access frequency (Section~\ref{sec:idx}). Under skewed data distribution, popular data items are concentrated into common cache lines using permutation indexes, leading to improved locality for query processing. By carefully organizing data at the cache line level, we can exploit the data skew for better performance, which has not be utilized by previous research.

The permutation index method is simple, yet very effective at improving the cache utilization. Figure~\ref{fig:distribution:cache} illustrates the cumulative frequency of cache line accesses for the example book dataset during query {\tt Q1}. Under a randomized data layout, the most popular book titles are sparsely distributed, so the distribution of cache line accesses is smoothed out. Even with significant data skew, the most frequent 8,000 cache lines (about the size of L1 cache using 4-byte {\tt bid}s) only correspond to 22\% of the fact table data (see the dashed line). With the permutation index, frequent data are concentrated and the percentage increases to 52\%. In Section~\ref{sec:idx:analysis}, we further analyze the improvement in cache hit rates.

Built on permutation indexes, we develop efficient database operators in the presence of skew. Permutation indexes provide frequency information about data items, allowing for threshold-based algorithms that execute different code paths for items with different degrees of skew. To take full advantage of modern architectures, our implementation makes use of single-instruction multiple-data (SIMD) instruction sets, multithreaded execution, and software prefetching (Section~\ref{sec:op}).
% Section~\ref{sec:op} discusses how we combine these techniques to speed up basic database operations.

Finally, we present a detailed experimental evaluation of our techniques in Section~\ref{sec:exp}. We conduct experiments on Intel Skylake and Xeon Phi processors, using both synthetic microbenchmarks and real data sets. Our results show that exploiting skew and reordering data can significantly improve performance, making queries using permutation indexes up to an order of magnitude faster.
% depending on the platform and the degree of skew in data.

\section{Background}
\label{sec:bg}

\subsection{Data Skew}
\label{sec:bg:skew}

As introduced in Section~\ref{sec:intro}, a wide variety of real-world phenomena approximately follow skewed distributions~\cite{clauset2009power, newman2005power}. For database systems, skewed distribution of empirical data has been recoginized in previous studies as a serious challenge to efficient query processing, since databases have to carefully handle parallel query execution and avoid performance degradation due to the long tail effect~\cite{martens2001classification}. Various partitioning algorithms and load balancing strategies have been studied to combat skew~\cite{dewitt1992practical,pavlo2012skew}. Different from these studies, we demonstrate that uneven distribution of data can in fact improve performance when the skewed accesses are exploited.
% , when our proposed techniques are employed to exploit skewed access patterns.
%  and effectively improve cache untilization.

While the techniques described in this paper apply to any skewed distribution, for ease of exposition, we focus on Zipf distributions in our microbenchmark study because they are common and allow one to model the degree of skew with a single parameter $z$. In a Zipf distribution, the frequency of the data item having rank $r$ is proportional to $r^{-z}$. $z=0$ corresponds to a uniform distribution, while many real-world skewed data sets can be modeled by Zipf distributions with $z \approx 1$~\cite{clauset2009power, moreno2016large}.

\subsection{Data Representation}
\label{sec:bg:rep}

For physical storage, we assume a columnar format as is common in analytic databases. An individual column is represented as a dense array of integers. The compact integer representations can be used as data values and/or as foreign key offsets into dimension tables stored as a collection of column arrays. When the true data values are strings, products, or other types of non-numerical data, databases will typically use some form of dictionary coding to map the domain to a contiguous range of integers.

\subsection{Architectural Issues}
\label{sec:bg:arch}

We assume a columnar format as is common in analytic databases. An individual column is represented as a dense array of integers. The compact integer representations can be used as data values and/or as foreign key offsets into dimension tables stored as a collection of column arrays.
In order to perform well on such RAM-resident data, database management systems must be sensitive to modern computer architectures. We highlight two significant trends in modern architectures that influence implementation choices.

{\bf Single Instruction Multiple Data.} Modern processors support single-instruction multiple-data (SIMD) instruction sets. These instructions process many data items at a time, enhancing the data-parallelism of algorithms that can be written in a SIMD fashion. Further, SIMD instructions convert control dependencies to data dependencies, helping to eliminate branch misprediction latencies~\cite{zhou2002implementing}. Current mainstream CPUs (e.g., Skylake~\cite{doweck2017inside}) and Xeon Phi processors (Knights Landing~\cite{sodani2016knights}) now support the 512-bit extensions (AVX-512).

{\bf Cache and TLB Performance.} Modern architectures provide multiple levels of data and instruction caches. These small but fast memories improve performance when algorithms display spatial or temporal locality. Conversely, algorithms that ignore the caches (e.g., by randomly accessing data structures whose footprint exceeds the cache size) incur many cache misses and/or TLB misses that can reduce performance by an order of magnitude. Addressing cache and TLB performance is a necessity for a database management system given that most data structures for representing the underlying data will be much larger than the data caches. When access patterns are deterministic, prefetching can hide some of the memory access latency. For common patterns such as sequential access, the hardware can automatically prefetch data that is soon needed.

In some cases, these architectural features interact. For example, SIMD gather and scatter instructions are used to access multiple memory locations in a single instruction. Such instructions are highly efficient on L1-cache-resident data, but perform no better than scalar code on data outside the L1 cache~\cite{menon2017relaxed, polychroniou2015rethinking}. This kind of observation argues for synergy. SIMD optimizations enhance the potential of data locality to improve performance, and cache locality optimizations stand to benefit from additional SIMD enhancement.

\subsection{Cache Performance Optimizations}
\label{sec:bg:opt}

For the example queries of Section~\ref{sec:intro}, some existing techniques could be used to improve performance by optimizing the cache behavior. One approach is to measure the footprint of the accessed data column to see whether it is larger than the performance-critical cache.  If so, a range-partitioning pass over the input {\tt bid} references (and their fact table payloads for {\tt Q1} and {\tt Q2}) can redistribute the data into fragments. With a sufficient partitioning factor, each fragment will reference a subset of {\tt bid}s that fits into the cache. As long as the partitioning pass is done efficiently with mostly-sequential data accesses and few cache misses~\cite{polychroniou2014comprehensive}, the overhead of partitioning may be smaller than the gain from avoiding cache misses. On the other hand, multiple passes through the very large sales table would be needed, meaning that the overhead is nontrivial even if there is a performance improvement.

An alternative approach is to use software prefetching to overlap the latencies of multiple cache misses~\cite{menon2017relaxed}. A prefetching distance $d$ is determined empirically, and the {\tt bid} value for the record $d$ steps ahead of the current record is prefetched into the cache using machine-specific prefetch instructions. The hope is that the cache miss latency is paid while the processor is doing useful work on other items. Modern CPUs can have as many as 10 outstanding memory requests per core, allowing many prefetches to be in flight at once.

While prefetching helps by overlapping latencies, it does not completely eliminate cache miss effects in workloads that are memory-bandwidth bound. In the queries of Section~\ref{sec:intro}, there is limited work that can be done while waiting for the miss to resolve.
% For example, in {\tt Q1} the only non-load/store operations are a multiplication and an addition, which would typically take one or two cycles each, in contrast to the hundreds of cycles that could be needed to retrieve data from RAM.
Therefore, there remains an opportunity for methods that reduce the total volume of data that needs to be brought into the cache from slow memory.
\section{Permutation Indexes}
\label{sec:idx}

To address the cache utilization problem as outlined in Section~\ref{sec:intro}, we propose a novel index structure that we call a \emph{permutation index}. We start by identifying a fact table column $C$ with an integer data type. For simplicity we assume that the integer column is an offset into a dimension table as in the example queries. Let $|C|$ denote the number of distinct values appearing in column $C$.

\begin{figure*}
  \begin{subfigure}{0.32\textwidth}
    \centering
    \includegraphics[width=\textwidth]{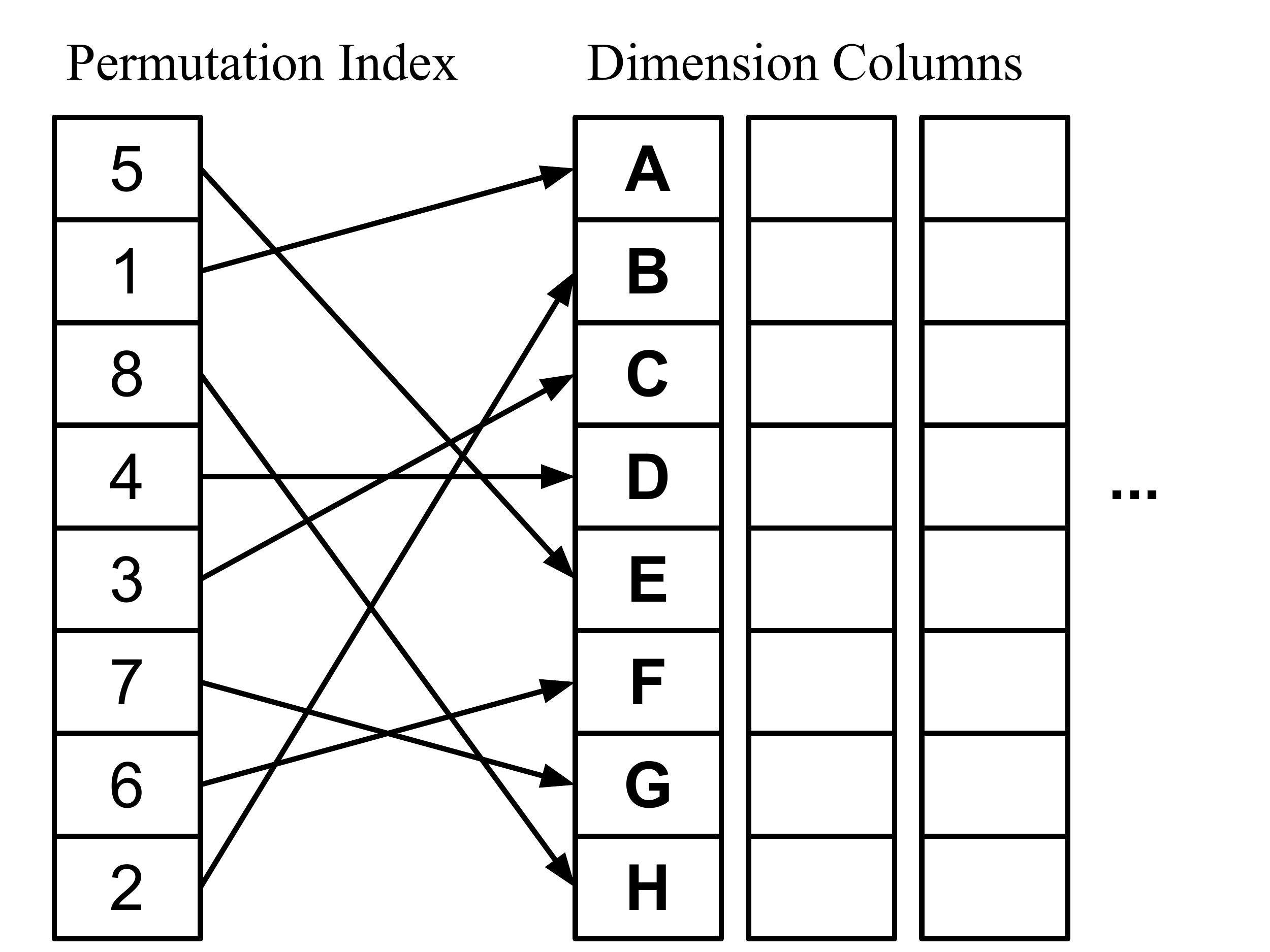}
    \caption{Dimension positions}
    \label{fig:idx:offset}
  \end{subfigure}
  \hfill
  \begin{subfigure}{0.32\textwidth}
    \centering
    \includegraphics[width=\textwidth]{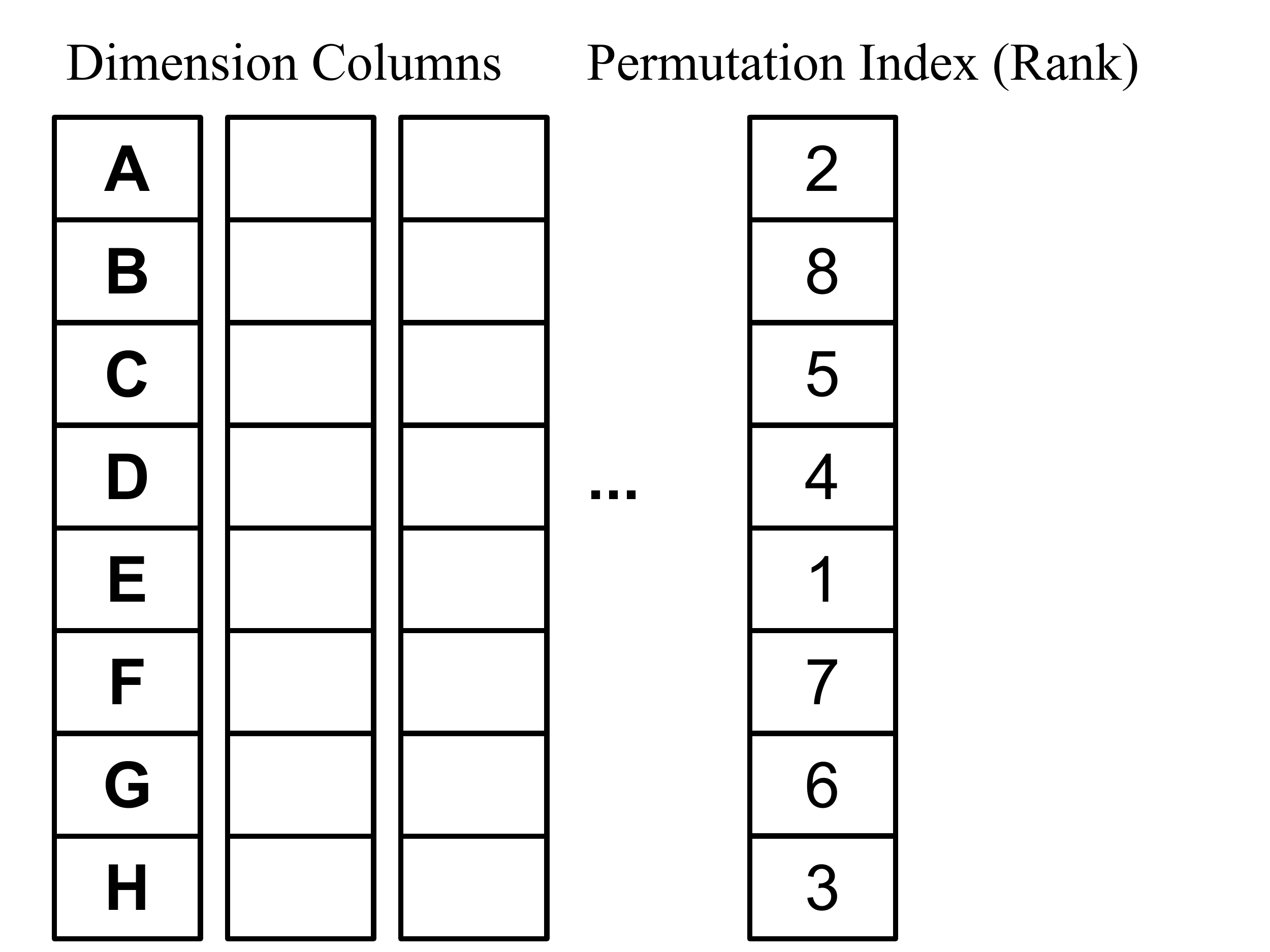}
    \caption{Frequency ranks}
    \label{fig:idx:rank}
  \end{subfigure}
  \hfill
  \begin{subfigure}{0.16\textwidth}
    \centering
    \includegraphics[width=\textwidth]{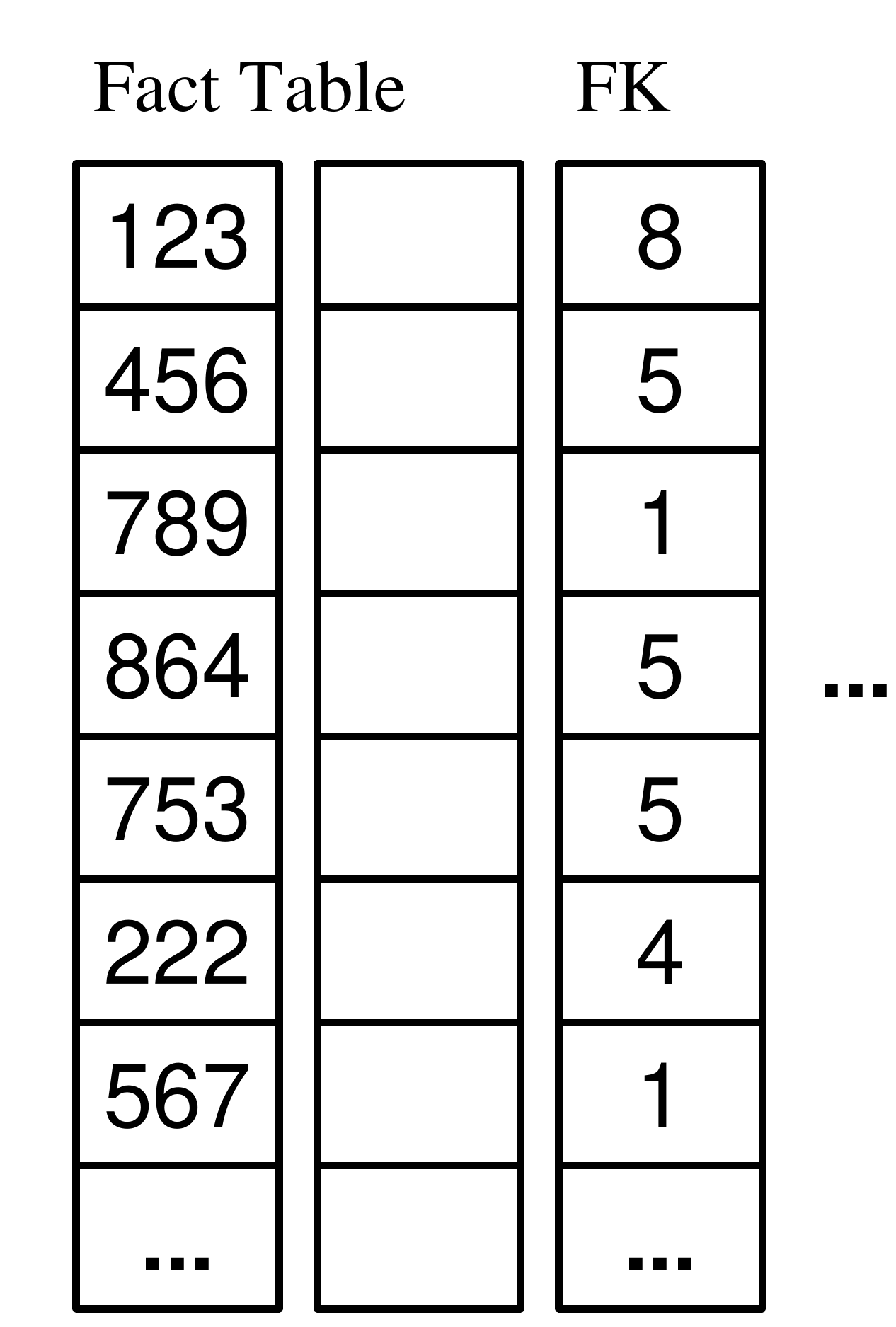}
    \caption{Original table}
    \label{fig:fact:original}
  \end{subfigure}
  \hfill
  \begin{subfigure}{0.16\textwidth}
    \centering
    \includegraphics[width=\textwidth]{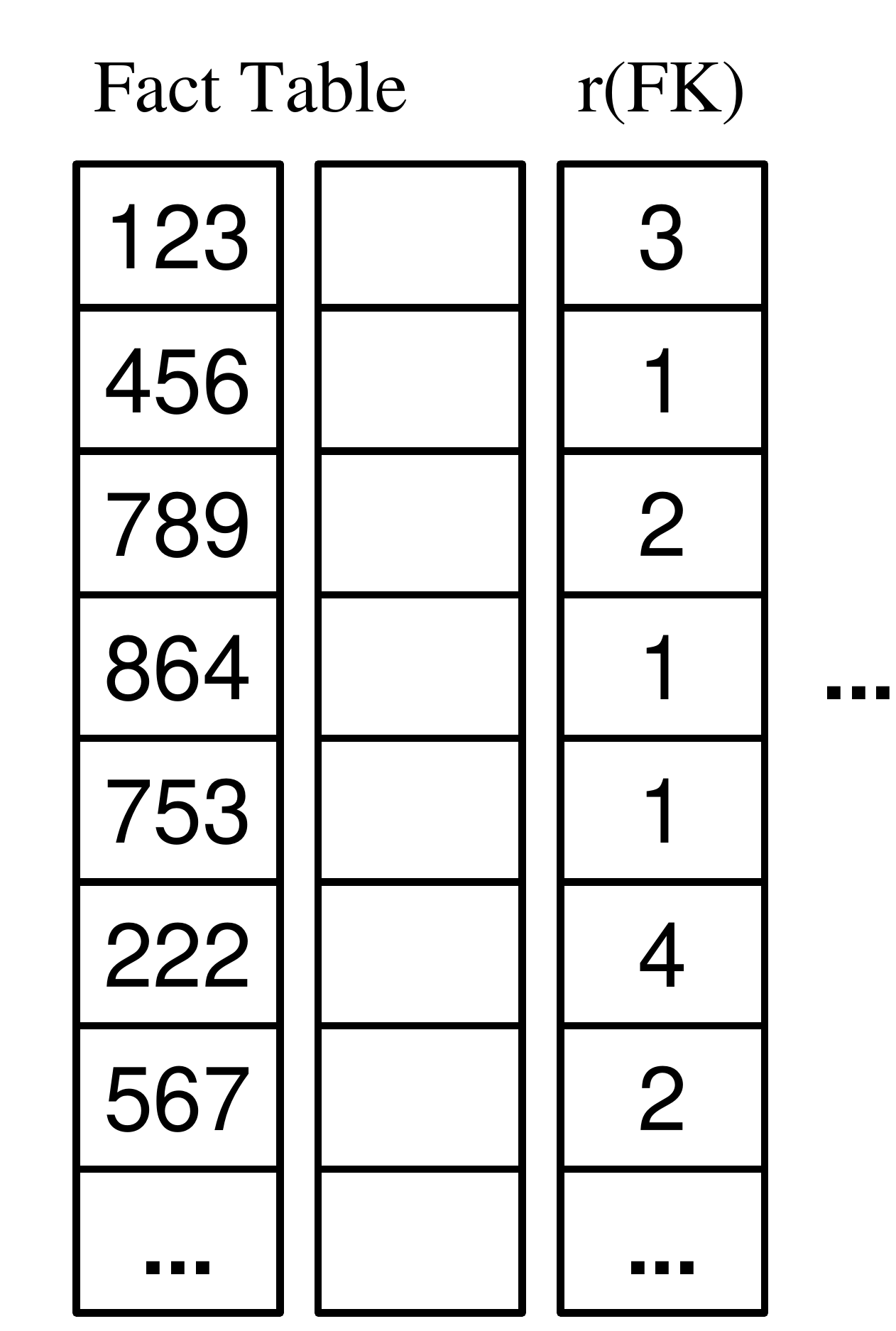}
    \caption{Transformed table}
    \label{fig:fact:transformed}
  \end{subfigure}
  \caption{(a) A permutation index, (b) its alternative representation, (c) a foreign key column $C$ before applying the permutation index transformation, and (d) the same column $C'$ with new identifiers after the transformation.}
  \label{fig:permutation_index}
\end{figure*}

The permutation index is an array of size $|C|$ containing the values appearing in column $C$ in decreasing frequency order. The left part of Figure~\ref{fig:idx:offset} shows an example in which 5 is the most frequent value in the referencing column (Figure~\ref{fig:fact:original}), then 1, 8, 4, etc. The arrows in Figure~\ref{fig:idx:offset} show how these integers may be interpreted as offsets into a dimension table (starting from 1). Figure~\ref{fig:idx:rank} shows an equivalent representation of the same information in which the popularity rank of each distinct data item in $C$ is stored as an additional column of the dimension table. If there are dimension table offsets that do not appear in $C$, then the representation of Figure~\ref{fig:idx:offset} will be slightly more compact than Figure~\ref{fig:idx:rank} because no information needs to be stored for the missing values.

Note that the array representation of a permutation index using $|C|\log|C|$ bits is already within $O(|C|)$ bits of the information theoretic bound of $\log(|C|!)$~\cite{munro2012succinct}. To support efficient inverse operations, we can store a permutation index using compressed data structures for permutations, so that both lookups and reverse-lookups take sublinear time~\cite{barbay2008succinct, munro2012succinct}. Alternatively, we can store both the offset and rank representations, so that lookups in either direction take only constant time.

\subsection{Index Building and Maintenance}

To achieve a better access pattern, we must recode the references in column $C$ into a new version with the permuted references. For the data of Figure~\ref{fig:permutation_index}, references to dimension table row $i$ are replaced by references to row ${Rank}(i)$ according to the rank representation in Figure~\ref{fig:idx:rank} to obtain Figure~\ref{fig:fact:transformed}. We assume that this replacement in the fact table has occurred prior to the query execution, typically when the database administrator (or automated physical design optimizer) decided to build the permutation index. Replacement of one integer value with another yields no net change in space for column $C$. The space cost of building the permutation index is one integer value per dimension table row, and is independent of the size of column $C$ of the (presumably large) referencing column in the fact table.

Building the permutation index can be implemented as an aggregation query ({\tt Q3}) to obtain the frequency of distinct dimension values, taking linear time, followed by sorting the values by their frequencies. The sorting step takes super-linear time but it executes over the much smaller dimension domain. Then, the identifier replacement in a fact table can be implemented as a materialization query ({\tt Q1}) using the permutation index we just built, again taking linear time. After this transformation, the old column $C$ (Figure~\ref{fig:fact:original}) is dropped, and replaced with a new column $C'$ with the new identifiers (Figure~\ref{fig:fact:transformed}).
% Reordering a dimension column is also a materialization during the preprocessing.

Updates to the permutation index can be handled without major reorganization by simply ignoring the effect of fact table insertions/deletions on the frequencies of existing domain values. When a new dimension table row is inserted, it is appended to the end of the dimension table, and its identifier is appended to the end of the permutation index as the new least-frequent item. A small number of updates is unlikely to dramatically change the relative ordering of popular value frequencies from a large fact table, so these simple choices will preserve most of the performance advantages of the structure even after a few updates. After many updates (e.g., a batch update in a data warehouse) the index should be rebuilt.

\subsection{Overview of Query Processing}

The permutation index functions as a reordering template for preprocessing the dimension table during the initial phases of query processing. We create a copy of the needed dimension table columns for a query, permuted according to the corresponding permutation index. For Figure~\ref{fig:idx:offset}, assuming we needed the first column, we would create an intermediate result in the order E,A,H,D,C,G,F,B. Because the cost of the query is likely to be dominated by the large fact table containing the source column $C$, preprocessing the dimension table in this way will be relatively fast. Note that a dimension table may have several permutation indexes that refer to it from different source columns, each with different orders. If the dimension table is referenced just from a single source, then its reordering can be done entirely ahead of query processing.

Having reordered the data, popular items are adjacent and therefore are likely to share cache lines with other popular items. Therefore, during query processing if we access the data via offsets in the reordered table, we will get much better utilization of all cache levels, particularly when there is skew in an otherwise large domain. As we shall discuss in detail in Section~\ref{sec:op}, several basic database operations can make use of a permutation index and the reordered data to improve efficiency (in different ways). Since a permutation index simply replaces a fact table column and adds a copy of the dimension column, there is minimal impact on other database queries.

% \subsection{Physical Design}

Using permutation indexes over skewed data essentially reorders data to change the memory access pattern during query execution. Physically reordering the dimension table by access frequency would eliminate the need for an explicit permutation index. However, we can impose only one ordering on the dimension table and there may be many competing demands (e.g., ordering by a domain value to support lexicographic comparisions via identifiers, or additional references from fact columns having different skew properties). A similar observation holds for the fact table, where reordering rows might help locality in the ordered columns, but there can be only one order unless data is stored redundantly. In light of these observations, one can think of permutation indexes as a way to influence physical database design. For some skewed data columns, permutation indexes eliminate the need for physical ordering of the fact table by those columns in order to get good cache performance. Physical organization can then focus on the remaining columns that may have more serious nonlocality.

\subsection{Cache Behavior Analysis}
\label{sec:idx:analysis}

To decide whether to build a permutation index or not, we need to estimate the improvement in cache utilization. A sophisticated query optimizer also requires knowledge of the cache behavior to compute the cost of query execution, together with hardware characteristics. For these purposes, we now study a model for estimating the cache hit rates under different data layouts.

Prior work has developed analytical cache models based on stack distances~\cite{cabetacaval2003estimating, beckmann2016modeling}. The stack distance is the number of distinct cache lines referred between two references to the same line. For an LRU cache with $S$ lines, accesses with stack distance less than $S$ will be cache hits, while others are misses. To apply this model in our setting, we need an estimation of access frequency to the cache lines, and the stack distance distribution for each line.

Conventional databases store statistics about column distributions to help estimate the selectivities of query conditions. It is also possible to sample the data and fit the samples to known distributions~\cite{alstott2014powerlaw}. Most power-law distributions can be modeled with a few parameters, such as the slope and intercept in a log-log plot. Given the value frequency distribution, we can then map it to the cache line distribution for a particular layout. For example, using permutation indexes over 4-byte values, the access frequency of the most frequent 64-byte cache line would be the sum of the 16 largest frequencies in the value distribution. For a randomized mapping from value distribution to cache line distribution, the computed frequency would be an approximation.

Let $f_i$ denote the access frequency to cache line $L_i$, where $0 \le f_i \le 1$ and $\sum_i f_i=1$. For line $L_i$, we model its next reference as a geometric distribution with probability $p=f_i$. Suppose after $k$ trials, we see cache line $L_i$. According to the stack distance model, if there are fewer than $S$ distinct lines occurred within the $k$ trials, then the reference will be a cache hit. Due to data skew (espectially the very frequent lines), there are potentially repeated occurrences of the same cache lines within the $k$ trials. For every line $L_j$ where $j \ne i$, its expected number of occurrences is
$$n_j = k * f_j/(1-f_i)$$
If $n_j>1$, then there are $(n_j-1)$ repeated occurrences of $L_j$. Thus, an estimation of the number of distinct lines is
$$d = k - \sum_{j \ne i} \max(0, n_j-1)$$
As $k$ increases, $d$ also increases (at a lower rate).

For cache line $L_i$, we want to find a threshold $K$ so that after $K$ trials, we see $L_i$ and there are estimated $d=S$ distinct lines $L_{j \ne i}$ seen within the $K$ trials. Then we know for any $k<K$, we have $d<S$ so the access will be a cache hit. To compute the threshold $K$, a binary search can be used where $K$ is at least $S$. Modeled as a geometric distribution, the cumulative distribution (CDF) of $(k<K)$ for $L_i$ is
$$cdf_i = 1-(1-f_i)^K$$
Therefore, the overall estimated cache hit rate is $\sum_i (f_i \cdot cdf_i)$.

Empirically we find this model is accurate with errors less than 5\% on our microbenchmarks given that we have a good estimation of the cache line frequencies. In practice, we can also sample the cache line distribution directly to verify the accuracy of the model and adjust accordingly if there is partial clustering, which we plan to address as future work.

\section{Skew-Aware Operator Implementation}
\label{sec:op}

We now discuss how to implement efficient database operators that can take advantage of permutation indexes and the reordered data. As introduced in Section~\ref{sec:intro}, we focus on three types of basic database operations: materialization ({\tt Q1}), selection ({\tt Q2}), and aggregation ({\tt Q3} and {\tt Q3a}).

The benefits of the proposed permutation index approach are twofold. First, reordering data using permutation indexes improves cache utilization during execution of important database operations (Section~\ref{sec:idx}). Second, the transformed identifiers in the fact table (Figure~\ref{fig:fact:transformed}) provide valuable information about data frequency, which the operators can exploit to perform threshold-based processing (Section~\ref{sec:op:threshold}). In this way, the operators are ``aware'' of the degree of skew, and are able to take appropriate code paths for different actions.

We find that even a straightforward scalar implementation of these operators can take advantage of the improved cache utilization. To further enhance performance, we study the use of a variety of techniques including SIMD vectorization (Section~\ref{sec:op:simd}), multithreaded execution (Section~\ref{sec:op:multithread}), and software prefetching (Section~\ref{sec:exp:micro}).

\subsection{Data-Parallel Execution}
\label{sec:op:simd}

Data-parallel execution using SIMD instructions has been successful at speeding up various database operations~\cite{zhou2002implementing, kim2009sort, willhalm2009simd, satish2010fast, polychroniou2015rethinking}, especially when the data is cache resident.
% Here we study how we can combine SIMD execution with cache utilization optimization using permutation indexes.
% In our implementation, we use the new 512-bit extension (i.e., AVX-512), which doubles the register width in comparison to the previous AVX2. Although the core of AVX-512 is supported by both Intel's Skylake CPUs and Xeon Phi (Knights Landing) processors, the two types of processors also implement other AVX-512 extensions, resulting in slightly different implementations on different platforms. For example, Skylake CPUs support the AVX-512 DQ extension, enabling some enhanced integer operations, while Xeon Phi processors support AVX-512 PF, enabling advanced prefetch instructions for gather and scatter operations ({\tt vpgatterpf} and {\tt vpscatterpf}, see more details in Section~\ref{sec:exp:micro}).
For data-parallel read and write, we use AVX-512 gather and scatter instructions extensively.
% We also make use of some new AVX-512 instructions as explained shortly.
For simplicity of presentation, we assume a database column is simply an array of 32-bit integers. Most AVX-512 instructions provide variants for other data types including 8, 16, and 64-bit integers as well.

{\bf Materialization.} The operator uses the fact table values as indexes to perform gather instructions from the dimension data array. Given 32-bit integer values, an AVX-512 gather instruction retrieves 16 dimension values at once. The latency of this gather operation depends directly on the number of cache misses occurred during its execution (Section~\ref{sec:op:threshold:materialize}).

{\bf Selection.} This operation produces an array of qualifying fact table offsets. After preprocessing the dimension table using the selection condition to obtain a bitmap, the operator checks each row of the fact table against the bitmap using the referencing identifier as the index, writing out the row offset in the fact table if the bitmap testing succeeds (the scalar implementation needs to be branch-free to avoid the branch misprediction penalty). Whenever the operator needs to test a random bit, reads from the bitmap incur random memory accesses, similar to materialization. For SIMD, we use gather and shift instructions to compute addresses in the bitmap and to extract the bits into a mask. Using the AVX-512 compressed store instruction {\tt vpcompressd}, we can contiguously store the selected fact table offsets (those with their respective bits set in the mask) into an output array.

{\bf Aggregation.} This operation generates an array of numeric types to compute an aggregate of some fact table column, grouped by the dimension table offset. The implementation basically scatters into an output array after gathering old aggregates and peforming arithmetic computations. In a scalar implementation, the memory access pattern is similar to materialization with additional writes.

SIMD aggregation, however, needs an additional step to check for conflicts before scattering, since different SIMD lanes may write to the same memory location. When internal conflicts occur, we identify a subset of conflict-free SIMD lanes using an AVX-512 conflict detection instruction {\tt vpconflictd}, and allow only writes from this subset to succeed using the masked scatter instruction. The other SIMD lanes are retained for processing during the next iteration, alongside new data. If there are many conflicts (e.g., when fact table data is highly skewed), then performance deteriorates severely due to this conflict resolution step. In the worst case where all SIMD lanes attempt to update the same aggregate value, only one (instead of 16 assuming 4-byte integers) can proceed. We shall discuss this problem further in Section~\ref{sec:op:threshold:aggregate}. Note that these SIMD implementations do not rely on permutation indexes. Data-parallel optimizations and cache locality optimizations are orthogonal, but they work together to further boost performance.

% Note that these SIMD implementations do not rely on permutation indexes. Data-parallel optimizations and cache locality optimizations are orthogonal, but they work together to further boost performance as mentioned in Section~\ref{sec:bg:arch}. Next we study how permutation indexes can be used to improve current operator implementations.

\subsection{Threshold-Based Processing}
\label{sec:op:threshold}

An important insight in the use of permutation indexes is that the transformed identifier in the fact table (see Figure~\ref{fig:fact:transformed}) can be used as a proxy for the value frequency in the table. Given an estimate of the function mapping the numeric identifier to the value frequency, the system can estimate the likely cache behavior at each level using the model in Section~\ref{sec:idx:analysis}. In general, the most frequently accessed data are likely cached under common cache replacement policies. Given the available cache capacity, it is possible to derive a threshold $t$ such that accesses to data items with identifiers smaller than $t$ are likely cache hits, while identifiers larger than $t$ lead to misses. We can therefore choose different code paths for items based on their anticipated cache residence using a simple comparison between the identifiers and the threshold $t$.

\subsubsection{Materialization}
\label{sec:op:threshold:materialize}

\begin{figure}
  \centering
  \includegraphics[width=0.33\textwidth]{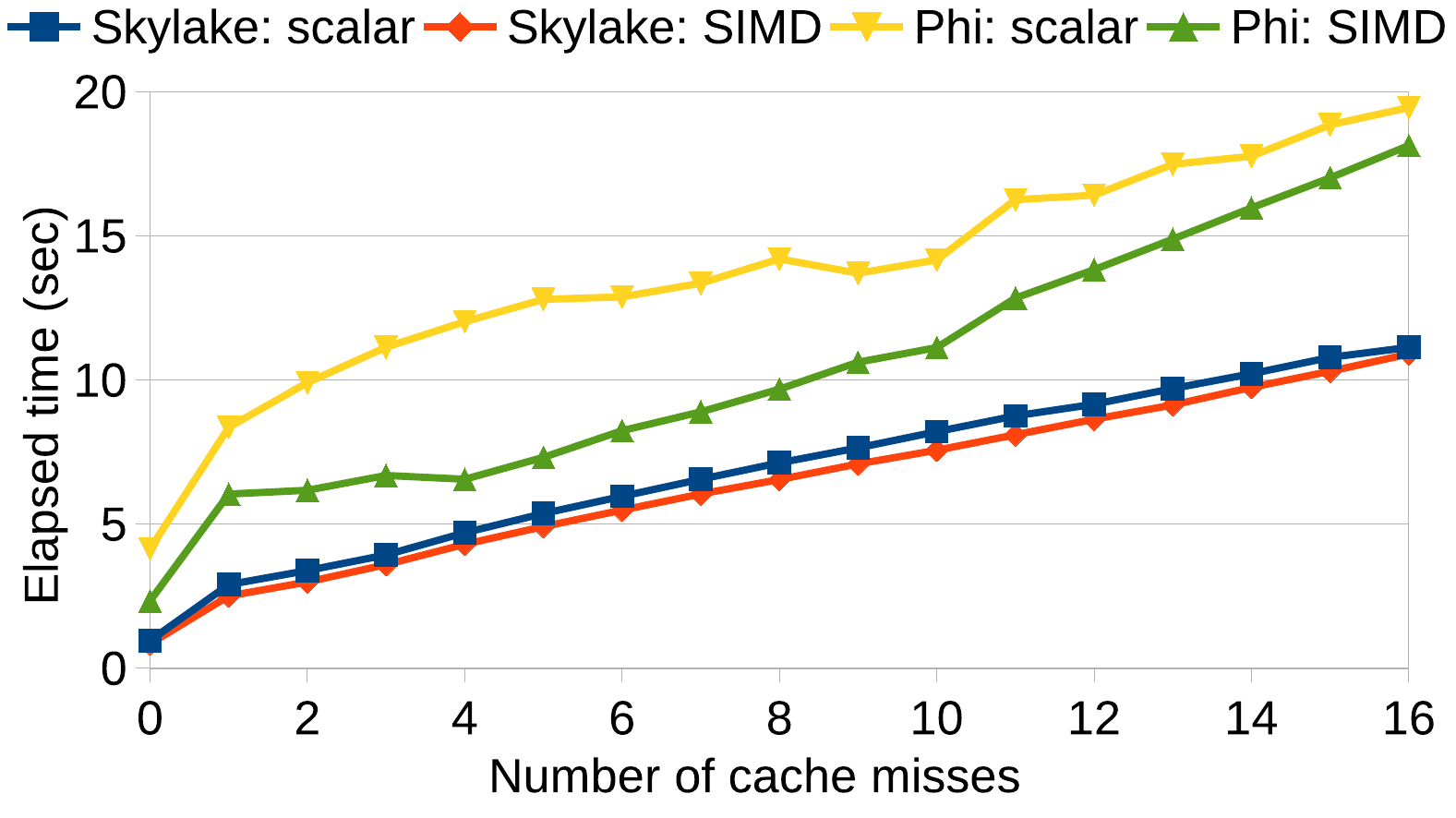}
  \caption{Materialization performance with varying cache misses per 16 elements.}
  \label{fig:materialize_cache}
\end{figure}

One situation where such an optimization pays off is when most, but not all data references are cache hits. Consider a materialization query like {\tt Q1} in which roughly one in 16 accesses is a cache miss, and the other 15 accesses are cache hits. A profile of the performance of materialization, with data constructed to achieve specific cache miss rates, is shown in Figure~\ref{fig:materialize_cache}. For this profiling, we use 128 million dimension keys and 1 billion fact table tuples (the default microbenchmark setting in Section~\ref{sec:exp:micro}). Using single-thread implementations of both scalar and SIMD versions described in Section~\ref{sec:op:simd}, we measured the elapsed time of the query execution on a Skylake processor and a Xeon Phi processor.

On both platforms, there is a big jump from 0 misses to 1 miss per 16 references, and then a less steep increase with additional misses. When there are no misses, the code can run at full capacity with essentially no memory traffic. At one miss, there is a stall on average once for each SIMD instruction. Additional misses have a less dramatic impact because the system can have multiple outstanding misses at any given time.
% Note that unlike the previous generation, the Knights Landing cores~\cite{sodani2016knights} are super-scalar with out-of-order execution similar to mainstream CPUs.

Another observation from Figure~\ref{fig:materialize_cache} is that the SIMD version has about 2x better performance over the scalar version when data is cache resident for the Phi. When there are many cache misses and the performance is bounded by the memory bandwidth, scalar and SIMD implementations have similar performances. In other words, SIMD optimization is most effective for accessing cache-resident data.

Given the apparent performance difference between in- and out-of-cache accesses, it would therefore be advantageous to split a data stream with a low cache miss rate into two pieces: The first (and largest) fragment always hits the cache and runs at cache speeds. The smaller fragment contains the likely misses, but when misses occur they occur together and their latencies can be overlapped.

The implementation for such a splitting method can be done in a data-parallel way. Based on an estimated threshold, a SIMD comparison operator determines which items are likely hits, and their references are resolved immediately. We use masked gather instructions in AVX-512 to directly write materialized data into the output array. For items failing the comparison, the item and its address are written to the tail of a buffer. A second pass then iterates through the buffer to resolve the remaining references. In Section~\ref{sec:exp:micro}, we demonstrate that for a materialization operation, the splitting method performs better than the baseline when there are one or two cache misses per 16 accesses.

\subsubsection{Aggregation}
\label{sec:op:threshold:aggregate}

The frequency information implicit in transformed identifiers can also be used to improve the performance of aggregation. As we discussed in Section~\ref{sec:op:simd}, SIMD implementations of aggregation perform poorly when the degree of skew is high in data. The reason for this performance degradation is that the most popular data items become so common that conflicts within the SIMD scatter step are frequent. In other words, two or more different SIMD lanes try to update the aggregate value (e.g., count) for the same group. Conflicts take several steps to resolve, and during this resolution process additional conflicting values can be read into the SIMD register, exacerbating the problem.

To mitigate this behavior, one can avoid conflicts on common items by re-mapping accesses to the most frequent items to distinct copies, one per SIMD lane. Identifying whether an item is among the most common is simple: it has an identifier that is no more than some threshold $t$. Because of data skew, remapping just a few of the most frequent items can already be quite effective in reducing conflicts. For example, in our microbenchmark experiments (Section~\ref{sec:exp:micro}), remapping the 40 most common 4-byte values was a good choice. In the modified aggregation algorithm, we keep 16 copies of each of the top 40 values. Accesses to these common values never conflict because the copy used is determined by the SIMD lane. The remaining items have only one copy, as before. The copies can be stored immediately before the original data array, preserving data contiguity. For SIMD lane $i$ (where $i=0 \ldots 15$), if the identifier loaded for this lane is not greater than $t$, then the copy at offset $(-i*t)$ is updated.

Using SIMD instructions, the remapping can be simply implemented as masked arithmetic computations based on the comparison with the threshold $t$. The top-$t$ locations in the original array now store only the aggregates from a single SIMD lane. When $t=40$, the cache footprint is bigger by $40*15*4=2400$ bytes, which is small relative to typical cache sizes and thus of minor impact. In the end, combining the copies for all SIMD lanes produces the final aggregates. In our experimental evaluations, we find this copying method effectively reduces conflicts in case of high skew, drastically improving performance.

{\bf Heavy hitter aggregation.} The permutation index has direct benefits for queries like {\tt Q3a} that compute counts for just the most frequent elements (i.e., heavy hitters). The most common 4,000 values for {\tt Q3a} are simply the transformed identifiers from 1 to 4,000, assuming that the permutation index is up-to-date. We can directly aggregate these values and ignore the rest (using the limit value as a cut-off threshold), unlike conventional methods that would need to compute the exact counts for all {\tt bid} values. The cache footprint of this approach is much smaller. For 4-byte integers, the output array is entirely L1-cache resident. Since there are almost no out-of-cache memory accesses, the performance of heavy hitter aggregation is significantly better than a full aggregation.

\subsection{Intra-Query Parallelization}
\label{sec:op:multithread}

A modern CPU contains many cores, each capable of independent work. It is therefore essential that this available parallelism is exploited by implementing multi-threaded versions of query processing algorithms. Fortunately, materialization ({\tt Q1}) and selection ({\tt Q2}) operations are relatively easy to parallelize. We can partition the input fact data into non-overlapping chunks so that different threads can work on different chunks independently. Shared reads (e.g., from a common dimension array) typically do not cause performance issue since they are cached across cores.

Aggregation ({\tt Q3}) is trickier because independent threads may try to update the count for the same {\tt bid}. Similar to the SIMD conflict detection, implementations must guarantee the atomicity of potentially conflicting updates to the shared result array, which may induce a performance overhead. Data skew potentially exacerbates the problem by increasing the probability that conflicting updates occur.

On the other hand, to fully avoid conflicts and the overhead, each thread can have its own independent copy of the result array, so that threads do not concurrently update the same memory location. The obvious disadvantage is memory consumption, which is multiplied by the number of threads used. Another cost is the overhead of the final aggregation that combines partial results from all threads, which gradually becomes non-negligible with more threads used.
% Amdahl's law: given the same amount of data and increasing number of threads.

In case of data skew, we are able to get the best of both approaches. We use a hybrid approach to allow different threads access to their own private versions of the hot data, while using atomic operations on shared representations of the less frequent data. Such an approach reduces conflicts while remaining space-efficient, similar to recent approaches to parallel aggregation~\cite{cieslewicz2007adaptive, cieslewicz2010automatic, ye2011scalable}. Given the available memory space and the frequency estimator, it is possible to derive (or obtain experimentally) a threshold for distinguishing the frequent data based on the transformed identifiers. As we demonstrate in our experiments, a small threshold is often enough to speed up queries in the case of skewed data. For example, typically each thread can use a private buffer for the most frequent 8,192 data items so that the buffer fits in the L1 cache, using four-byte identifiers. A limitation of the current SIMD extensions is that they lack support for vector atomic instructions. Introduction of such instructions could significantly improve performance~\cite{kumar2008atomic}.
% code divergence between updating private and shared arrays.
% heavy hitter aggregation~\cite{polychroniou2013high}.

% In practice, there are some architecture and system issues one has to consider, espectially when comparing single-threaded and multithreaded performances. First, in multi-socket servers, RAM is organized in banks that are local to a socket. RAM from remote sockets can be accessed, but at a higher latency and lower bandwidth. This non-uniform memory access (NUMA) effect suggests that, to the extent possible, algorithms should operate on core-local data. In our implementation, we explicitly allocate buffers in local memory and pin threads to hardware cores. Second, for cache optimization, the last level cache (L2 on Phi, and L3 on Skylake) is shared in a multithreaded implementation, so the effective cache size per thread is smaller than an exclusive single-thread implemenation. Third, most modern CPUs reduce clock rates significantly when multiple threads are working \todo{avx-512 downclocking}.

\section{Experiments}
\label{sec:exp}

We conducted experiments on an Intel Skylake CPU and on an Intel Xeon Phi (Knights Landing) processor, both of which support AVX-512. The Xeon Phi does not have a L3 cache, and its 1MB L2 cache is shared between two adjacent cores. Table~\ref{tab:hardware} summarizes the details of our experimental platforms. Both machines are configured to use 2MB hugepages to avoid the TLB thrashing problem discussed in Section~\ref{sec:bg:arch}. 1GB hugepages are also available and result in similar performance. Both hugepage options lead to much better performance than using the regular 4KB pages.
% The Xeon Phi does not have a L3 cache, but it has 16 GB of multi-channel DRAM (MCDRAM) on-chip. On the Intel Xeon Phi 7210 CPU we used in our experiments, we measured the bandwidth of DDR4 DRAM at $\sim$70 GB/s and the bandwidth of the on-chip MCDRAM at $\sim$295 GB/s \todo{Phi config}. Note that the cores of a single many-core CPU do not access all regions of memory at the same speed, similarly to mainstream CPUs with multiple NUMA regions.

\begin{table}
  \centering
  \caption{Hardware Specifications}
  \label{tab:hardware}
  \begin{tabular}{|l|c|c|} \hline
    Microarchitecture & ~~~~Skylake~~~~ & Knights Landing \\  \hline
    Model Number      & 8175M           & Phi 7210        \\  \hline
    Clock Frequency   & 2.5 GHz         & 1.3 GHz         \\  \hline
    Cores x SMT       & 24 x 2          & 64 x 4          \\  \hline
    L1 Size / Core    & 32 KB           & 32 KB           \\  \hline
    L2 size / Core    & 1 MB            & 512 KB          \\  \hline
    L3 Size           & 33 MB           & -               \\  \hline
    Memory Bandwidth  & 60 GB/s         & 55 GB/s         \\  \hline
  \end{tabular}
\end{table}

Our code was compiled using GCC 7.3 with {\tt -O3} optimization and loop-unrolling enabled, and ran on 64-bit Linux operating systems. Performance counters such as cache misses and branch misses were obtained using the perf events interface. We employed thread pinning to avoid undesired thread migration, maximizing the utilization of private cache and local memory on multicore NUMA platforms.

\subsection{Microbenchmark Results}
\label{sec:exp:micro}

We have implemented microbenchmarks to evaluate the potential performance improvements enabled by the permutation index. We simulate skew in empirical data by generating zipf distributions with varying zipf parameters. For a particular zipf factor, we vary the number of keys in a dimension table (i.e., the cardinality of the domain). We fix the column cardinality to 1 billion tuples in a fact table, and assign key values (4-byte integers) from the dimension domain to the fact table column under the chosen zipf distribution, in random order. We also experiment with varying sizes of the dimension domain.

For the baseline approach, the tuple identifiers in a dimension table are randomized (denoted as ``rand'' in the figures). For the permutation index approach, the dimension column is sorted by frequency and the referencing column in a fact table is preprocessed to hold transformed identifiers (denoted as ``freq''). Using both scalar and SIMD (AVX-512) implementations, we measure the elapsed time for database operations discussed in Section~\ref{sec:op}. For a particular operation, all tuples in a column are executed in a tight loop. We also evaluate the approaches both with and without software prefetching of the data references.

\begin{figure*}
  \begin{subfigure}{0.33\textwidth}
    \centering
    \includegraphics[width=\textwidth]{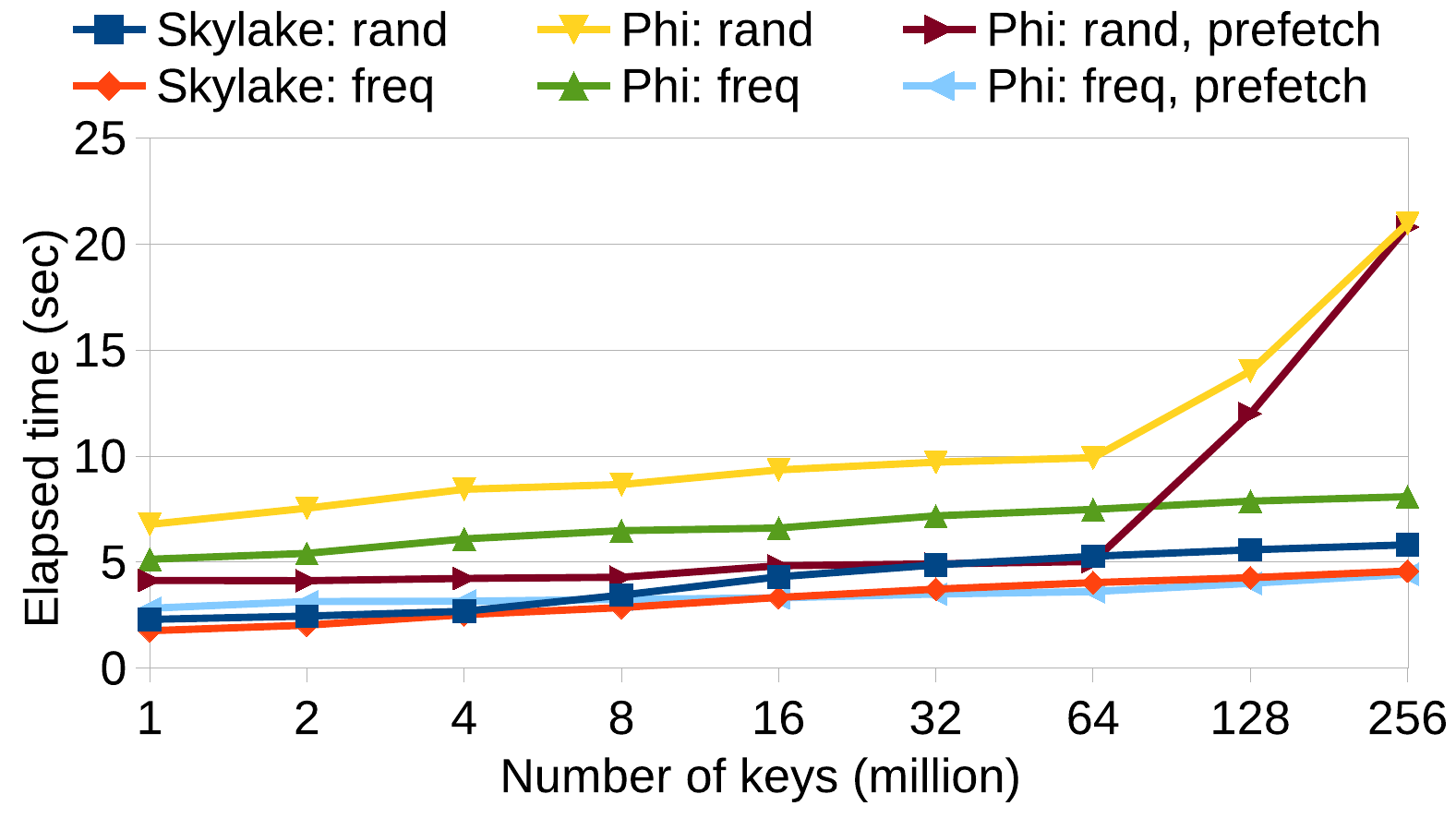}
    \caption{Varying number of keys, $z=1$\\~}
    \label{fig:exp:materialize:key}
  \end{subfigure}
  ~
  \begin{subfigure}{0.33\textwidth}
    \centering
    \includegraphics[width=\textwidth]{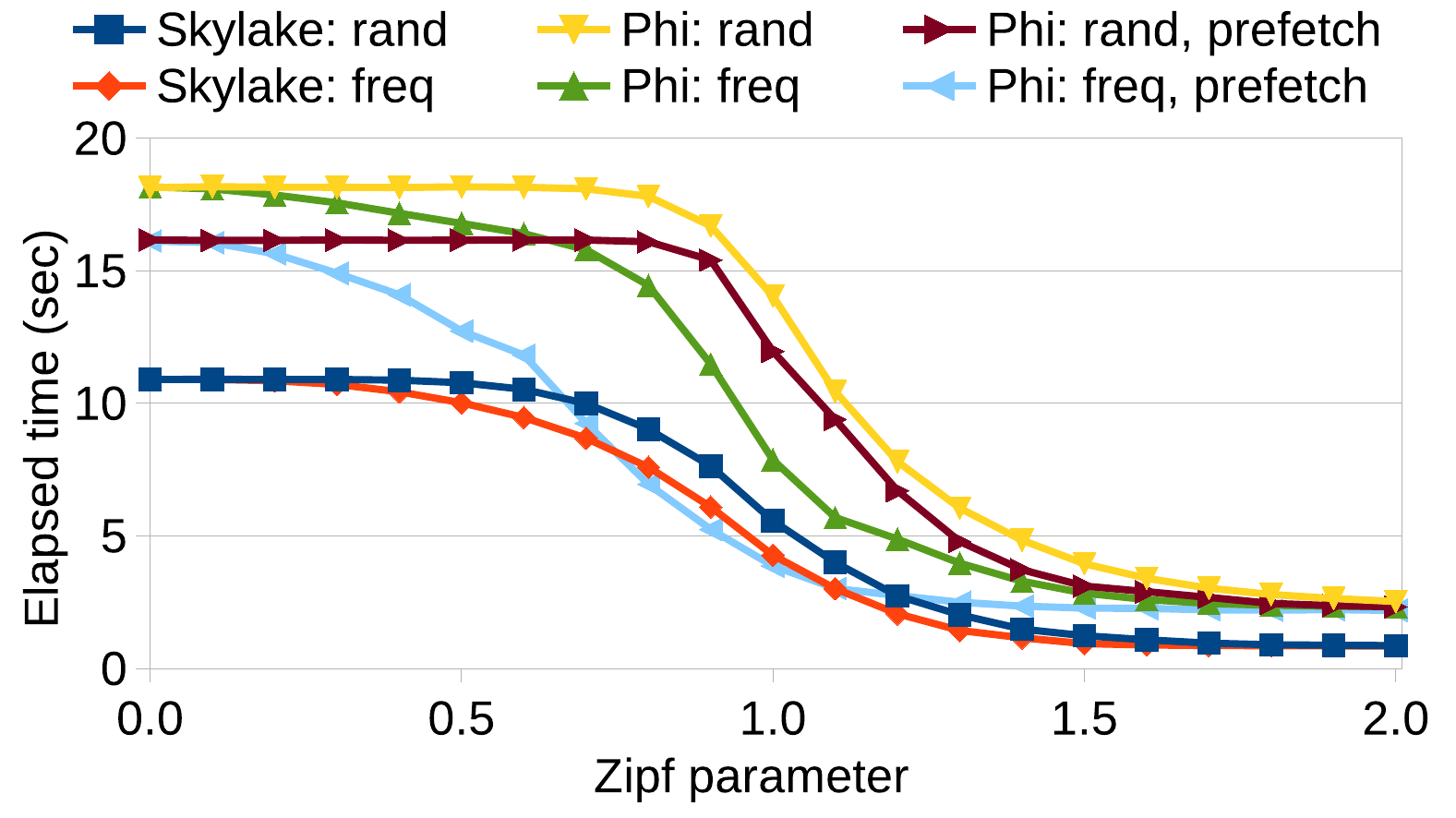}
    \caption{Varying zipf parameter, 128M keys\\~}
    \label{fig:exp:materialize:z}
  \end{subfigure}
  ~
  \begin{subfigure}{0.33\textwidth}
    \centering
    \includegraphics[width=\textwidth]{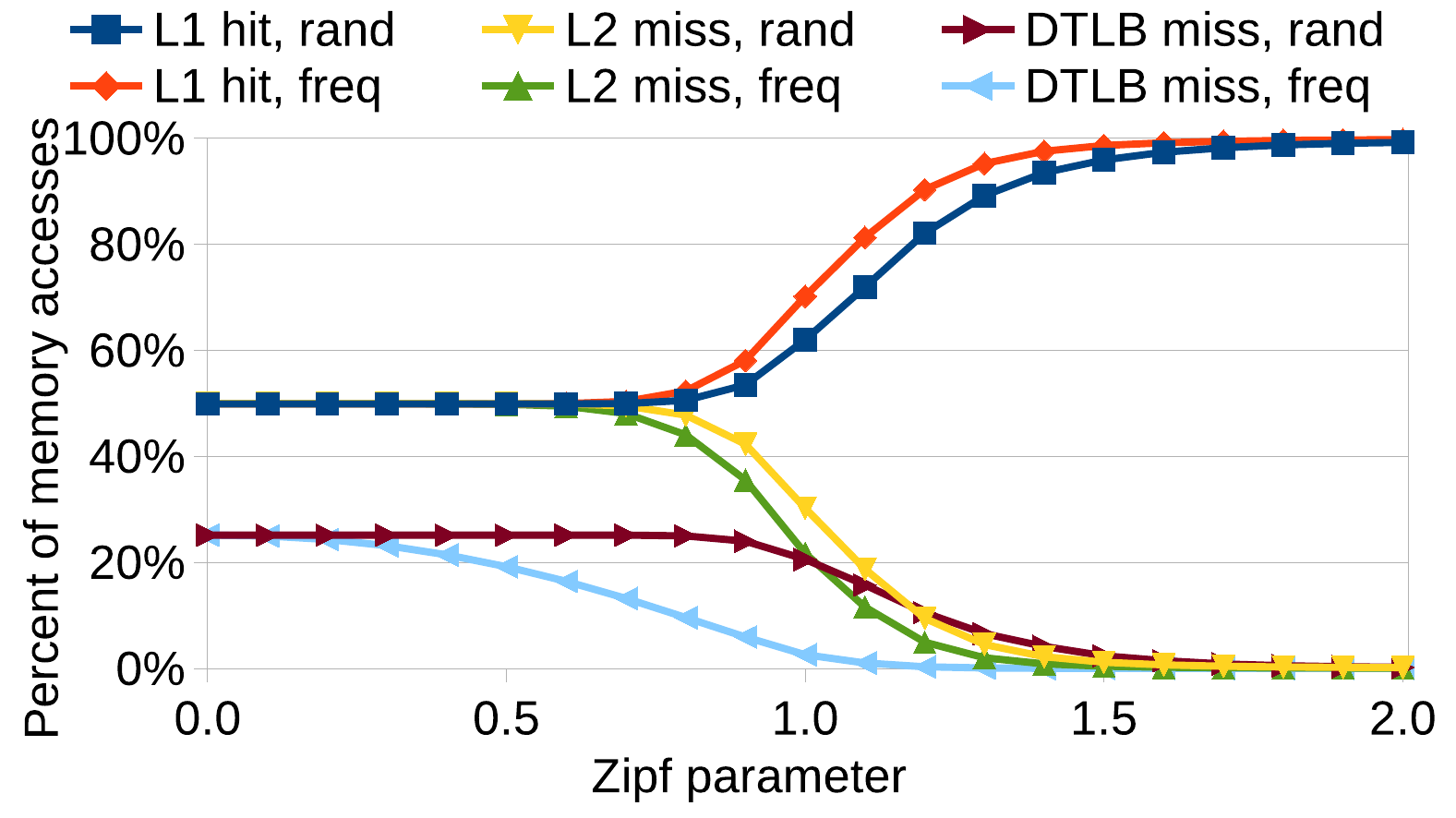}
    \caption{Cache behavior (Phi), 128M keys\\~}
    \label{fig:exp:materialize:cache}
  \end{subfigure}
  ~
  \begin{subfigure}{0.33\textwidth}
    \centering
    \includegraphics[width=\textwidth]{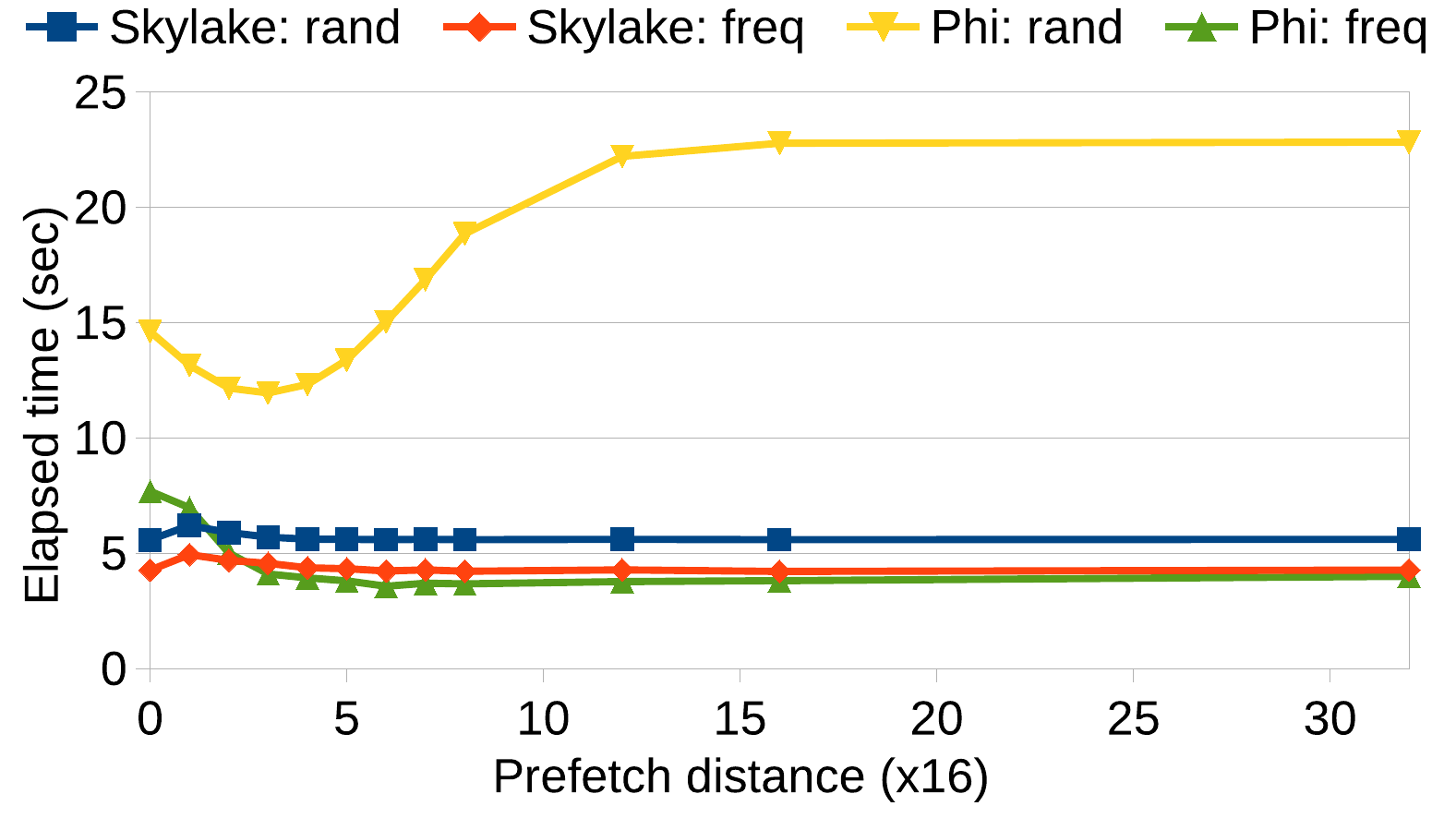}
    \caption{Varying prefetch distance, $z=1$, 128M keys}
    \label{fig:exp:materialize:prefetch}
  \end{subfigure}
  ~
  \begin{subfigure}{0.33\textwidth}
    \centering
    \includegraphics[width=\textwidth]{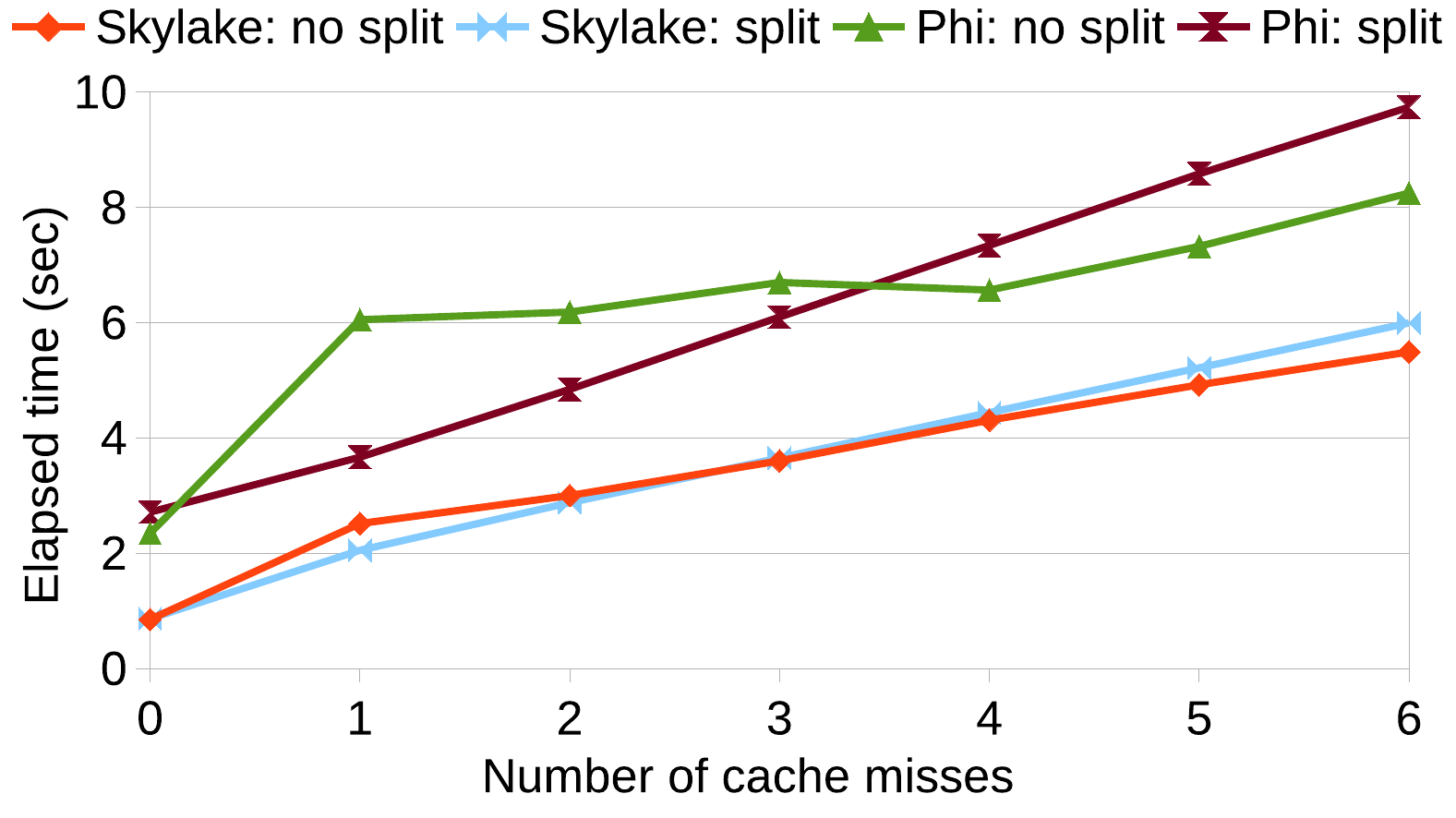}
    \caption{Varying cache misses per 16 elements, using threshold-based processing}
    \label{fig:exp:materialize:split}
  \end{subfigure}
  ~
  \begin{subfigure}{0.33\textwidth}
    \centering
    \includegraphics[width=\textwidth]{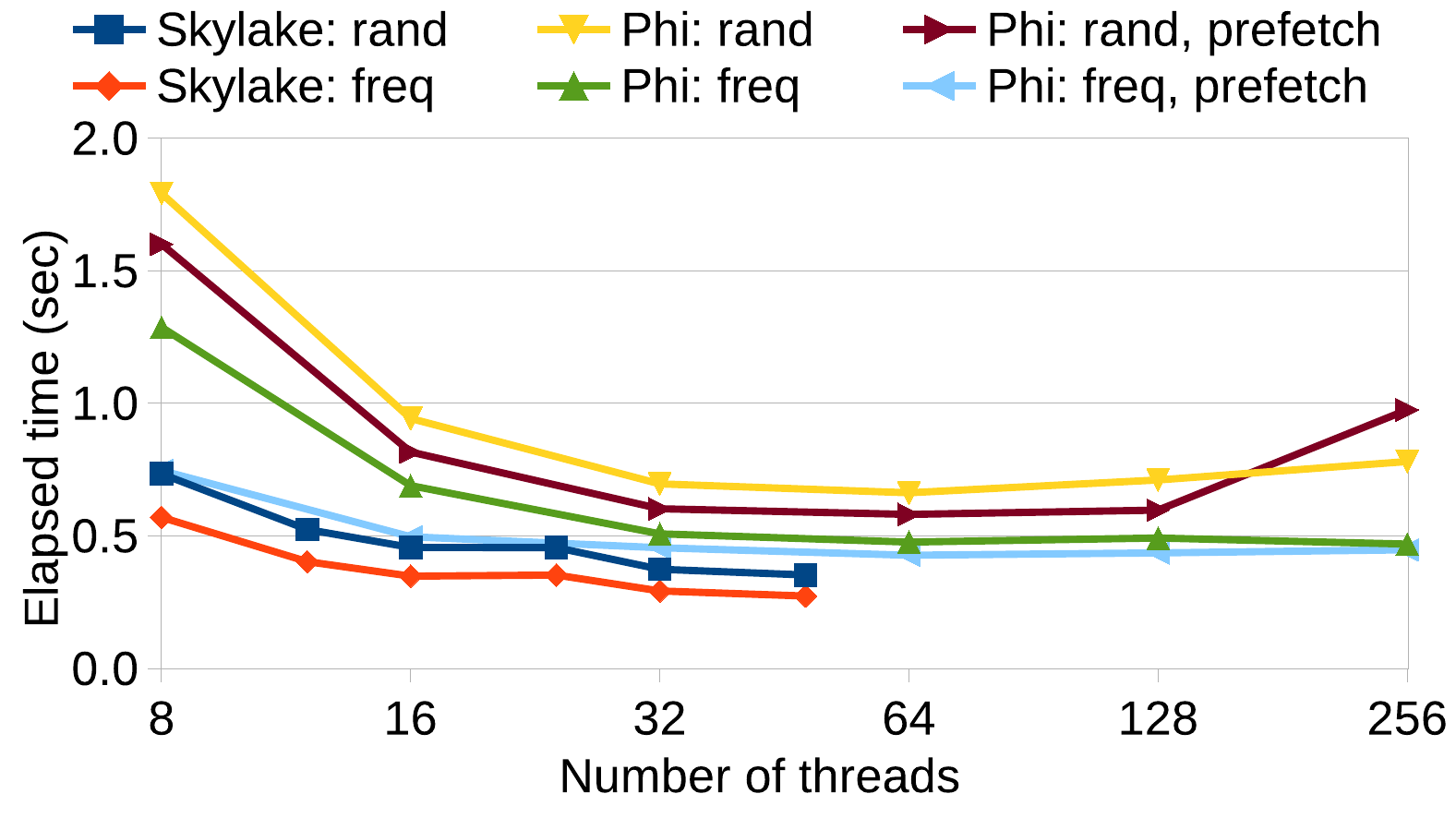}
    \caption{Multithreaded performance, $z=1$, 128M keys}
    \label{fig:exp:materialize:multithread}
  \end{subfigure}
  \caption{Materialization ({\tt Q1}) performance using permutation indexes (SIMD).}
  \label{fig:exp:materialize}
\end{figure*}

Figure~\ref{fig:exp:materialize} shows the performance of materialization ({\tt Q1}) using a SIMD implementation, which was faster than the scalar implementation for this query. Figure~\ref{fig:exp:materialize:key} shows that at $z=1.0$, a common distributional parameter in practice, permutation indexes speed up the query by more than 20\% on both architectures. With the number of keys fixed at 128 million, Figure~\ref{fig:exp:materialize:z} shows that performance improvements occur over a range of $z$ values. For smaller $z$ values, the data is close to uniform and both methods suffer cache misses. For larger $z$ values, the skew is so concentrated in a handful of data values that both methods enjoy cache hits most of the time. Without software prefetching, the speedup at $z=1$ and 128 million keys is 1.3x on the Skylake and 1.8x on the Phi.

Figure~\ref{fig:exp:materialize:cache} demonstrates that the time improvements on the Phi are indeed due to cache behavior. Half of the data loads are to the array of probes (almost always hits due to clustered sequential access and hardware prefetching), while half are to the accessed data in the dimension array. Thus, at small $z$ values we see a 50\% load hit rate in the L1 cache, and also a 50\% miss rate for the L2 cache. The dimension table footprint is much bigger than the L2 cache, making L2 hits that are L1 misses insignificant at small $z$ values. As $z$ increases, the L1 hit rate of the permutation index method is better (higher) than the randomized method, while the L2 miss rate corresponding to the access frequency from RAM is also better (lower). In addition, the permutation index method has much better data TLB utilization at 128 million keys. For example, at $z=1.0$, the DLTB miss rate is 2.6\% using permutation indexes, while the miss rate of the baseline method is 20\%.

% Note that at high $z$ values, prefetching could be deterimental to performance because most data accesses are already cache hits, so that prefetch instruction overheads accrue without changing the cache miss profile.
As discussed in Section~\ref{sec:bg:opt}, it is often possible to utilize out-of-order execution and software prefetching to hide the latency of cache misses. Figures~\ref{fig:exp:materialize:key} and \ref{fig:exp:materialize:z} also show that the benefits of permutation indexes are still present even when prefetching is used. For example, at $z=1.0$ and 128 million keys, using permutation indexes on top of prefetching makes the query 3x faster on the Phi. For these results, we chose the prefetch distances empirically. Figure~\ref{fig:exp:materialize:prefetch} shows performance with varying prefetch distances. Since Skylake does not support AVX-512 PF extensions, we simulate the {\tt vpgatherpf} instruction with a loop of 16 scalar prefetch instructions. As shown in the figure, the best prefetch distance on the Phi is 3(x16) for the baseline and 6(x16) for the permutation index method. As we prefetch more, the baseline approach quickly suffers due to cache thrashing, while the permutation index approach is almost unaffected. In general, we find that permutation indexes enable longer prefetch distances comparing with the baseline, making our approach even faster. On the Skylake, it turns out that the prefetch overhead is more than the benefits from prefetching, so we did not show its prefetch performance in Figures~\ref{fig:exp:materialize:key} and \ref{fig:exp:materialize:z}.

% Pattern doesn't matter.
In the case of high skew, Figure~\ref{fig:exp:materialize:split} presents a zoomed-in view of the SIMD performance (the red and green curves) from Figure~\ref{fig:materialize_cache} for 0--6 cache misses, together with the performance of the threshold-based splitting methods described in Section~\ref{sec:op:threshold:materialize}. Without splitting, there is a big difference of 2.1--3.1x from no cache misses at all to just a single miss, while subsequent misses cause much less latency. When the number of cache misses is small (1 or 2 per 16 elements), the threshold-based implementation splitting in- and out-of-cache accesses is better than the baseline. With more misses, the benefits of this optimization are gradually outweighed by the overhead of writing out intermediate buffers.

Figure~\ref{fig:exp:materialize:multithread} shows the performance of multithreaded materialization. We vary the number of threads from 1 to 256 on the Phi, and to 48 on the Skylake. Since the performance scales almost linearly up to 16 threads, we omitted the results with fewer than 8 threads used. Even though threads are sharing cache resources, the permutation index method always performs better, achieving a speedup of over 1.3x on both platforms with varying threads. On the Phi, using more than 64 threads does not provide any more benefits than using the baseline approach, and prefetching makes the query even slower. In contrast, the permutation index approach still slightly improves performance.

\begin{figure*}%[t]
  \begin{subfigure}{0.33\textwidth}
    \centering
    \includegraphics[width=\textwidth]{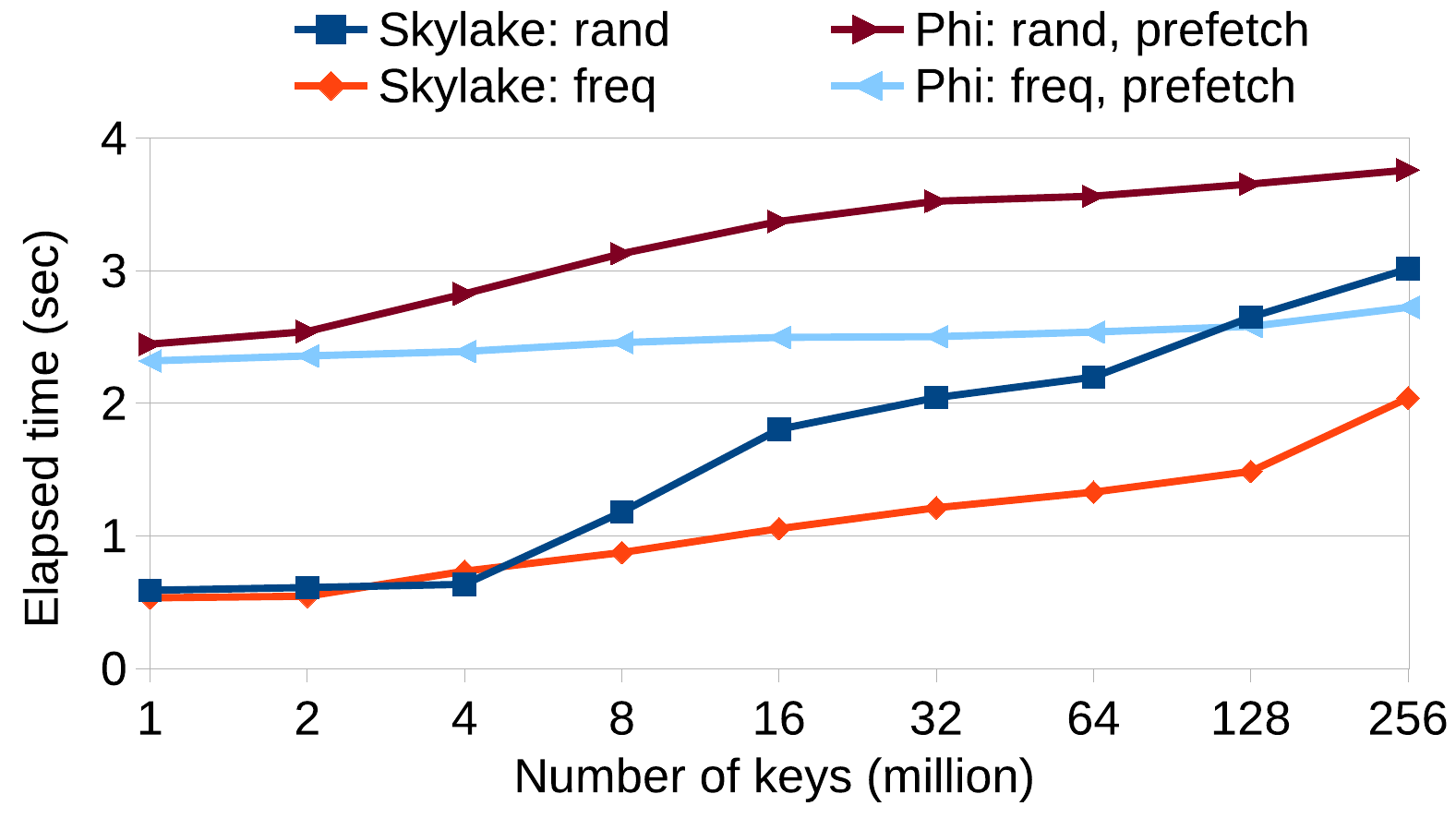}
    \caption{Varying number of keys, $z=1$}
    \label{fig:exp:select:key}
  \end{subfigure}
  ~
  \begin{subfigure}{0.33\textwidth}
    \centering
    \includegraphics[width=\textwidth]{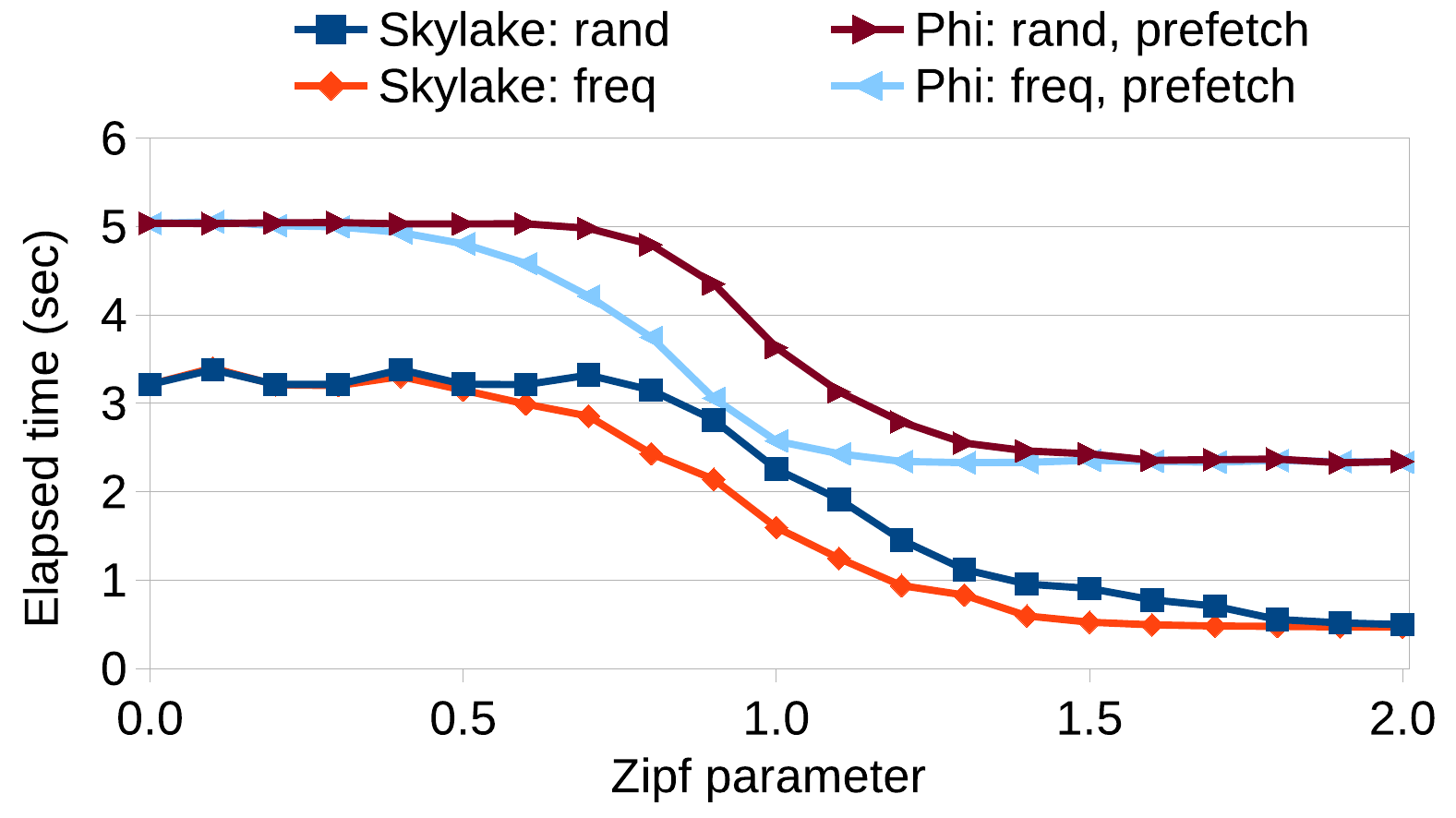}
    \caption{Varying zipf parameter, 128M keys}
    \label{fig:exp:select:z}
  \end{subfigure}
  ~
  \begin{subfigure}{0.33\textwidth}
    \centering
    \includegraphics[width=\textwidth]{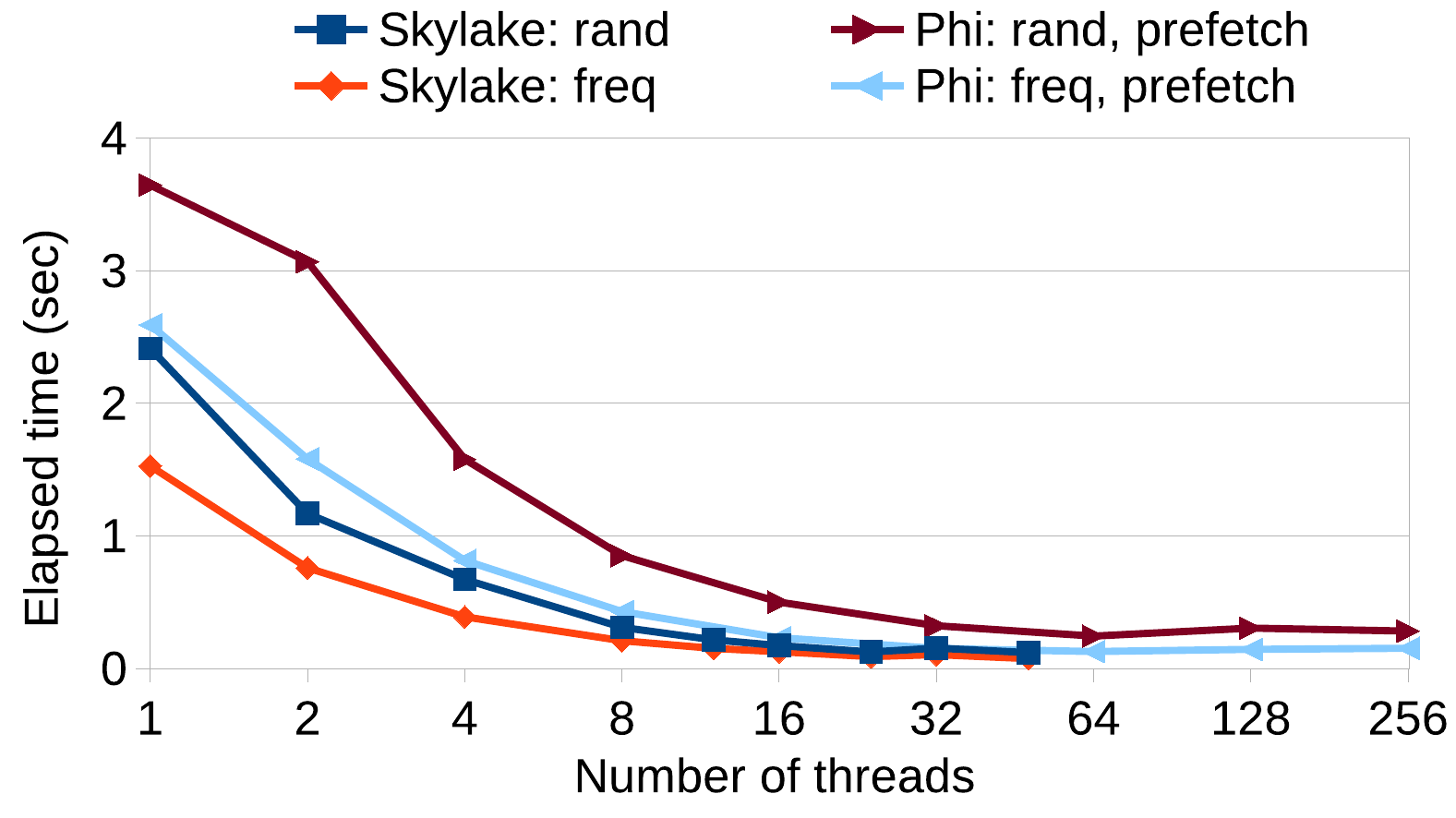}
    \caption{Multithreaded, $z=1$, 128M keys}
    \label{fig:exp:select:multithread}
  \end{subfigure}
  \caption{Selection ({\tt Q2}) performance using permutation indexes (SIMD).}
  \label{fig:exp:select}
\end{figure*}

Figure~\ref{fig:exp:select} shows the performance of selection ({\tt Q2}) using the SIMD implementation, which was again faster than the scalar implementation for this query. We omit the results without prefetching on the Phi since they are slower. The performance profile is similar to Figure~\ref{fig:exp:materialize}, except that for low key-counts the performance is better. Because only one bit (rather than 32 bits) is needed per key, a larger number of keys is required before the cache capacity is exceeded. The relative performance of the Skylake processor is better in Figure~\ref{fig:exp:select} than in Figure~\ref{fig:exp:materialize} because it has a large L3 cache that can hold the entire bitmap. Figure~\ref{fig:exp:select:multithread} shows the multithreaded performance. Comparing with the baseline, the permutation index method is 1.4--2.2x faster on the Phi, and 1.3--1.8x faster on the Skylake.

\begin{figure*}
  \begin{subfigure}{0.33\textwidth}
    \centering
    \includegraphics[width=\textwidth]{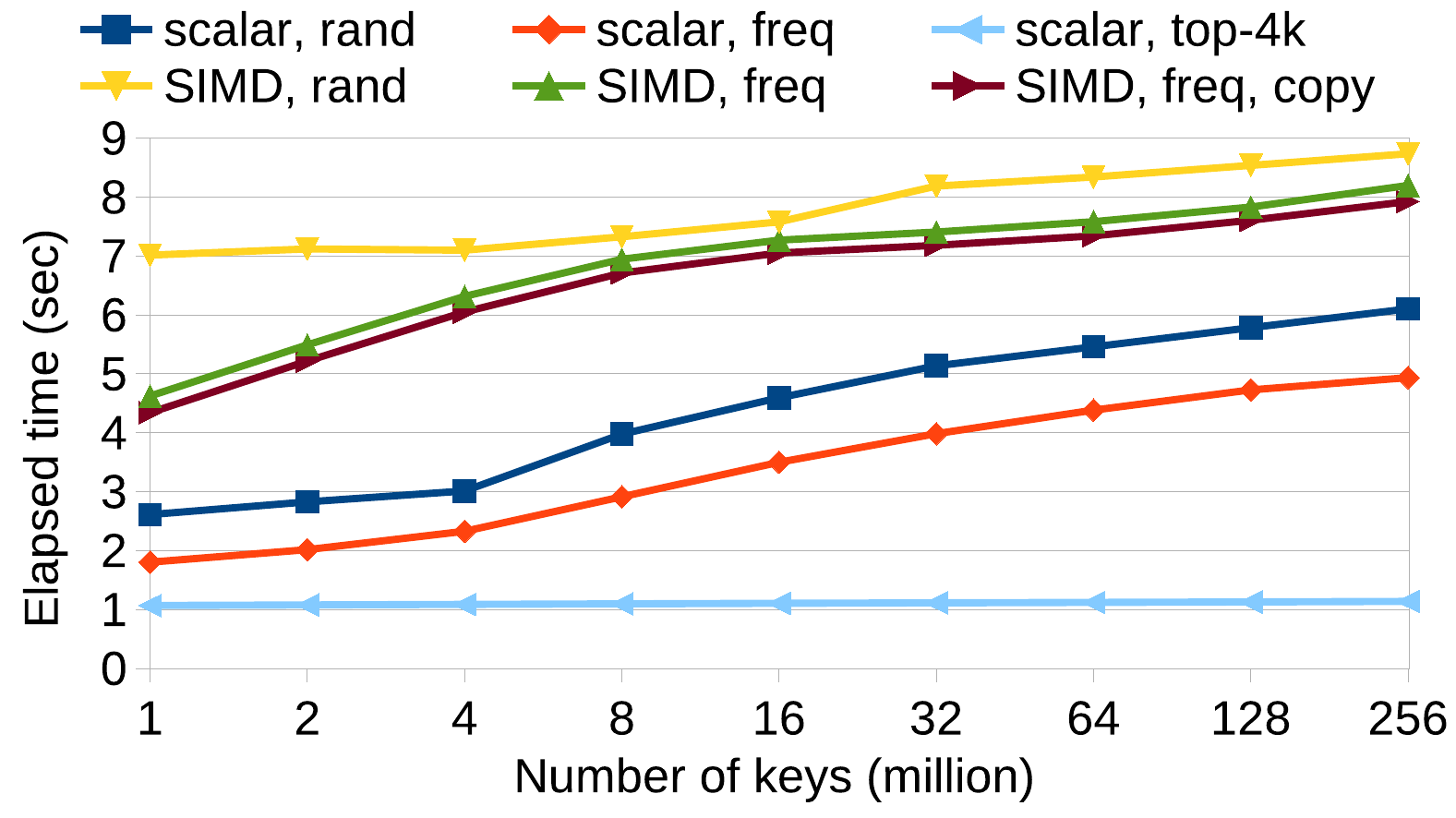}
    \caption{Skylake, varying number of keys, $z=1$}
    \label{fig:exp:aggregate:skylake:key}
  \end{subfigure}
  ~
  \begin{subfigure}{0.34\textwidth}
    \centering
    \includegraphics[width=\textwidth]{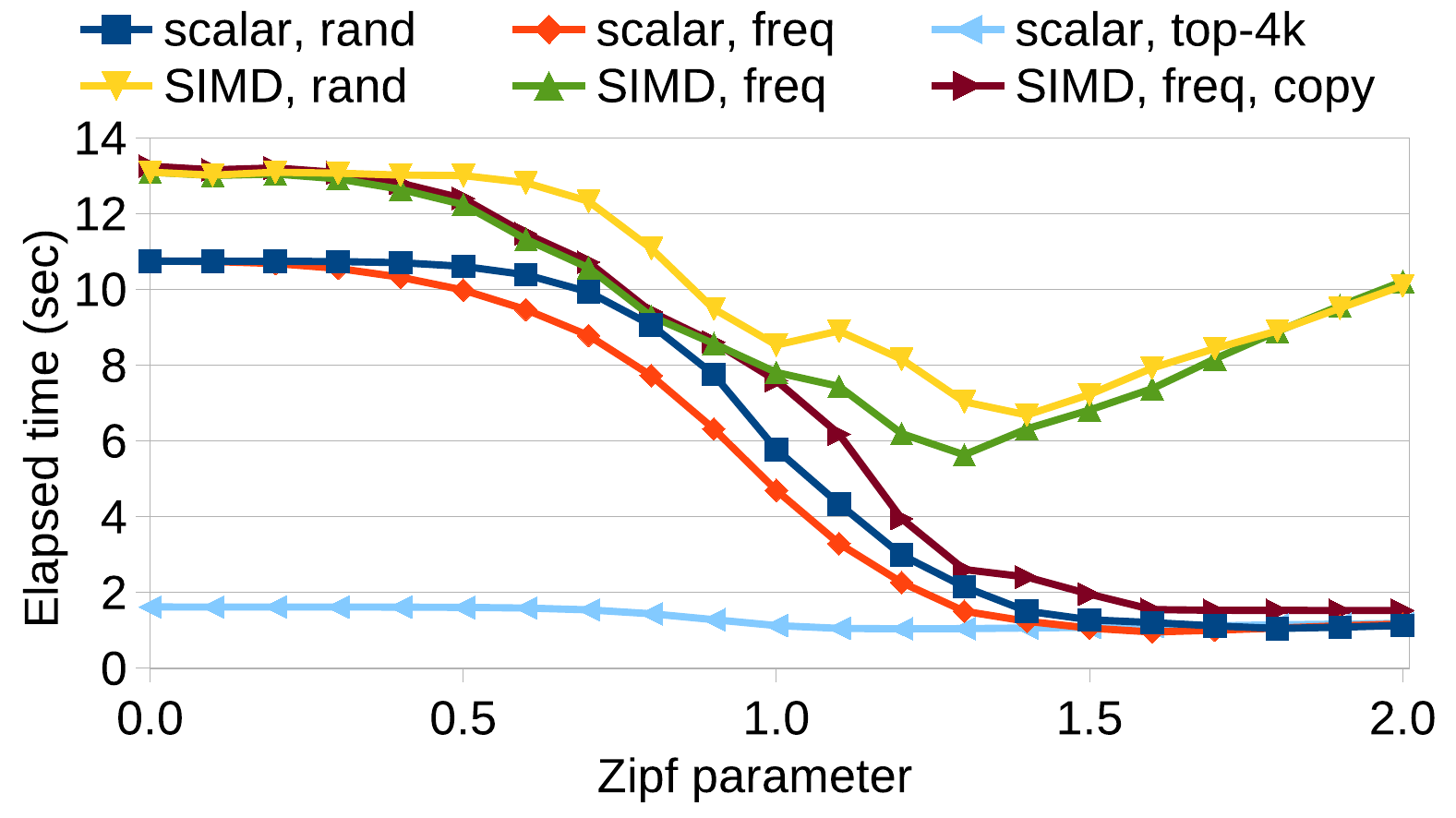}
    \caption{Skylake, varying zipf parameter, 128M keys}
    \label{fig:exp:aggregate:skylake:z}
  \end{subfigure}
  ~
  \begin{subfigure}{0.33\textwidth}
    \centering
    \includegraphics[width=\textwidth]{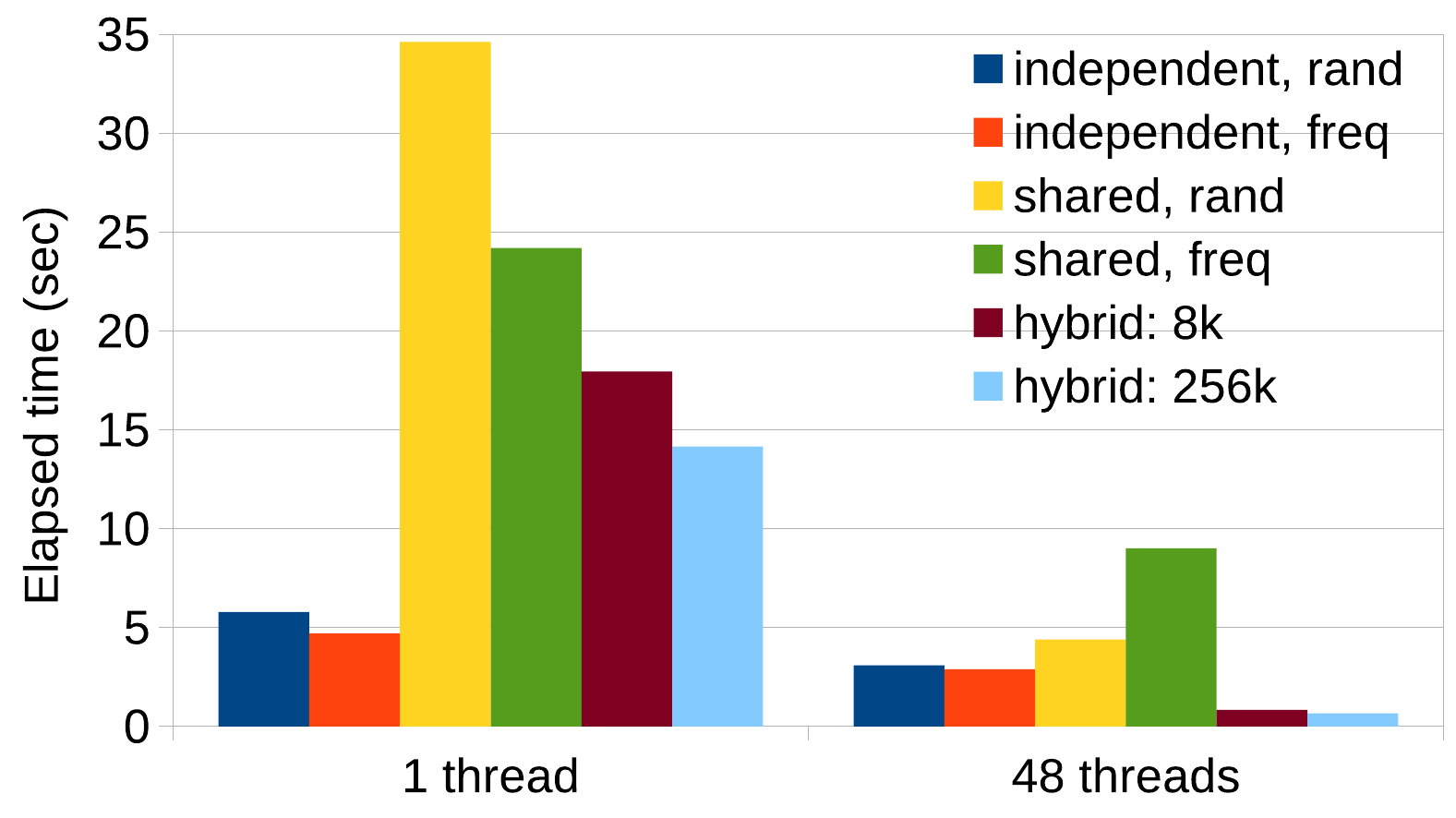}
    \caption{Skylake, multithreaded, $z=1$, 128M keys}
    \label{fig:exp:aggregate:skylake:multithread}
  \end{subfigure}
  ~
  \begin{subfigure}{0.33\textwidth}
    \centering
    \includegraphics[width=\textwidth]{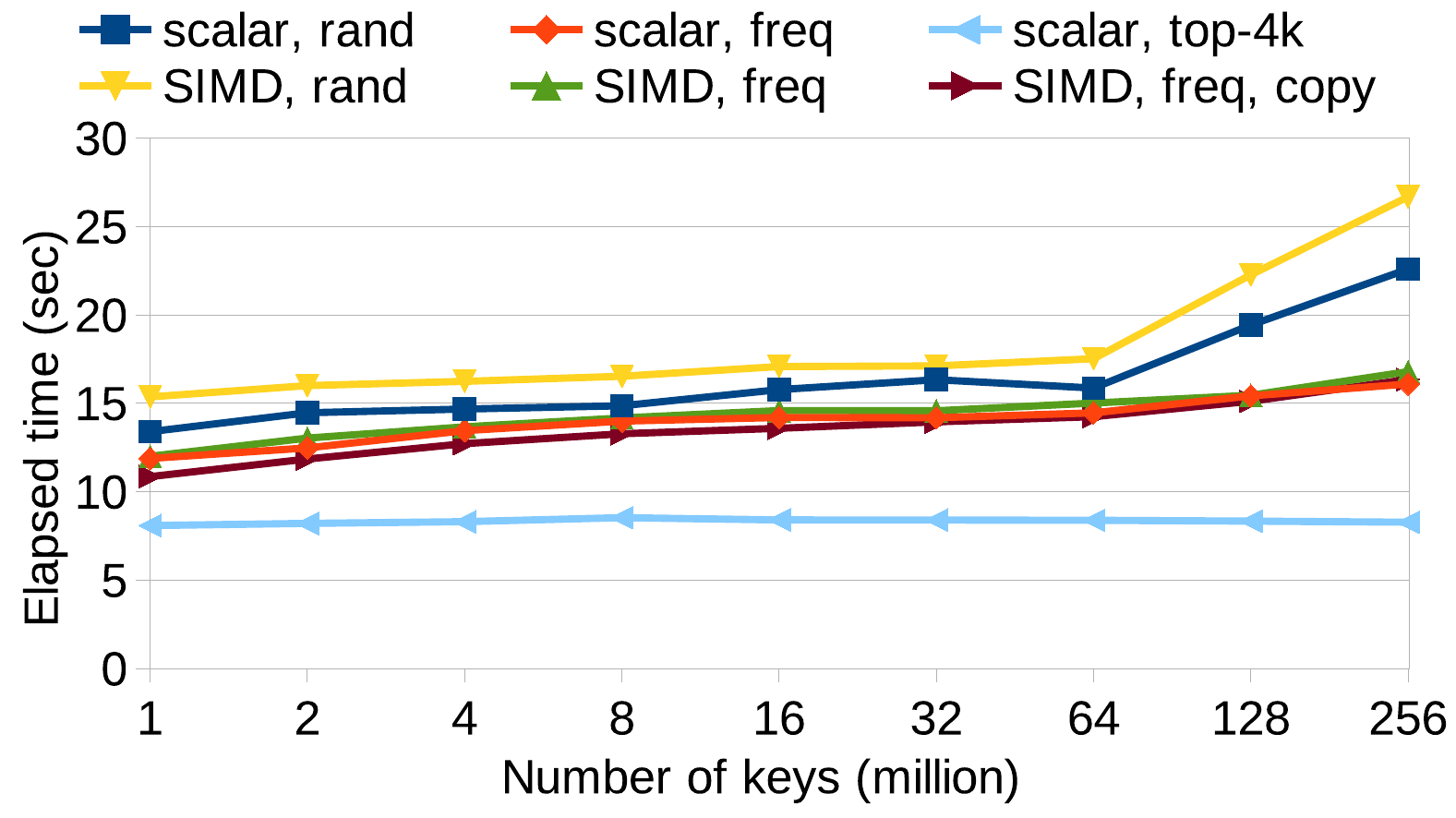}
    \caption{Phi, varying number of keys, $z=1$}
    \label{fig:exp:aggregate:phi:key}
  \end{subfigure}
  ~
  \begin{subfigure}{0.34\textwidth}
    \centering
    \includegraphics[width=\textwidth]{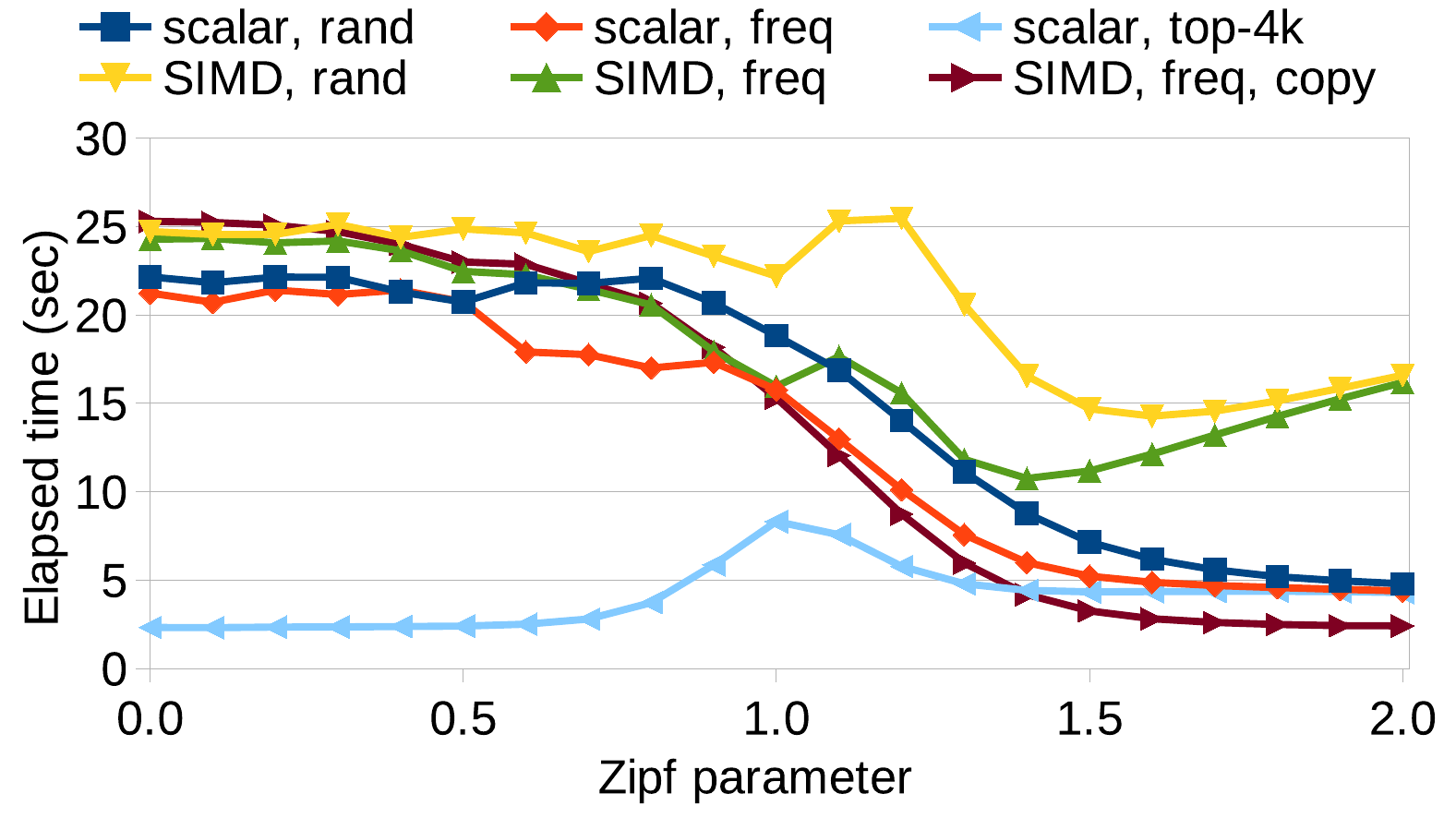}
    \caption{Phi, varying zipf parameter, 128M keys}
    \label{fig:exp:aggregate:phi:z}
  \end{subfigure}
  ~
  \begin{subfigure}{0.33\textwidth}
    \centering
    \includegraphics[width=\textwidth]{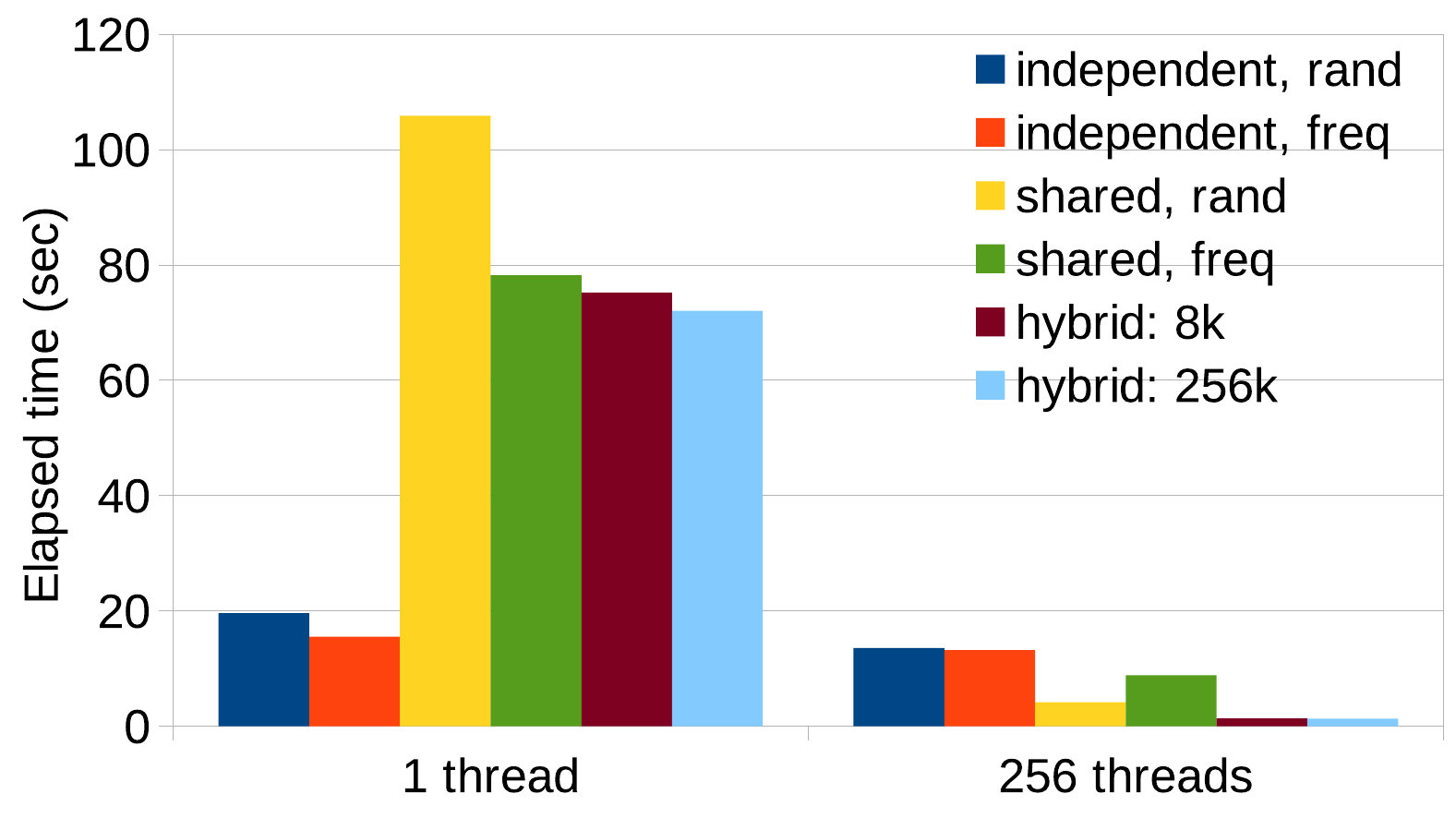}
    \caption{Phi, multithreaded, $z=1$, 128M keys}
    \label{fig:exp:aggregate:phi:multithread}
  \end{subfigure}
  \caption{Aggregation ({\tt Q3}) performance using permutation indexes.}
  \label{fig:exp:aggregate}
\end{figure*}

Figure~\ref{fig:exp:aggregate} shows the performance of aggregation ({\tt Q3}) using the scalar and SIMD implementations. On Skylake, scalar code generally outperforms the SIMD code because the SIMD code has the overhead of conflict detection and resolution. In fact, for high $z$ values, the impact of conflict resolution creates a severe degration in performance. The threshold-based copying method described in Section~\ref{sec:op:threshold:aggregate} addresses this issue. On Skylake, the scalar version with our permutation index optimization is fastest for all $z$ values in Figure~\ref{fig:exp:aggregate:skylake:z}. The modified SIMD algorithm with copying is the best-performing method at high skew levels on the Phi (Figure~\ref{fig:exp:aggregate:phi:z}), but not on the Skylake machine.

Figure~\ref{fig:exp:aggregate} also shows the performance of heavy hitter aggregation for computing the counts of the 4,000 most frequent data items (``top-4k''). Our optimization discussed in Section~\ref{sec:op:threshold:aggregate} significantly improves query performance, since the threshold-based approach only updates cache-resident data. In comparison, performing full aggregates without using the permutation index would have much larger memory footprint and have to do extra work to extract the top-4,000 results. As an example, for $z=0.5$, the performance improvement over computing the entire aggregate is 8.6x for the Phi, and 6.6x for the Skylake processor.

Note that we report the performance results for all operations using branch-free implementations, with one exception: for the heavy hitter aggregation on the Phi, we use an explicit if-branch to check whether the identifiers are smaller than 4,000, since we find the scalar branch-free version is particularly slow on the Phi. As shown in Figure~\ref{fig:exp:aggregate:phi:z}, when there are no heavy hitters ($z=0.0$), or when the data is very skewed ($z=2.0$), branching does not hurt the performance of heavy hitter aggregation. At high skew, the latency is longer because of more updates. When $z$ is close to 1.0, performance suffers due to branch misprediction penalty. For example, at $z=1.0$ the branch misprediction rate is 45.5\%. In contrast, Figure~\ref{fig:exp:aggregate:skylake:z} shows the performance of the branch-free heavy hitter aggregation on the Skylake, where there is no performance degration around $z=1.0$.

% \begin{figure*}
%   \begin{subfigure}{0.5\textwidth}
%     \centering
%     \includegraphics[width=\textwidth]{figures/aggregate_skylake_multithread}
%     \caption{Skylake}
%     \label{fig:exp:aggregate:skylake:multithread}
%   \end{subfigure}
%   ~
%   \begin{subfigure}{0.5\textwidth}
%     \centering
%     \includegraphics[width=\textwidth]{figures/aggregate_phi_multithread}
%     \caption{Phi}
%     \label{fig:exp:aggregate:phi:multithread}
%   \end{subfigure}
%   \caption{Multithreaded aggregation ({\tt Q3}) performance using permutation indexes, $z=1$, 128M keys.}
%   \label{fig:exp:aggregate:multithread}
% \end{figure*}

Figures~\ref{fig:exp:aggregate:skylake:multithread} and~\ref{fig:exp:aggregate:phi:multithread} present the aggregation performance using all 48 threads on the Skylake, and 256 threads on the Phi, respectively. Single-threaded results are shown as a reference. We compare the performance of independent-buffer implementations, shared buffer implementations using atomic operations, and the threshold-based hybrid approach described in Section~\ref{sec:op:multithread}. For independent-buffer implementations, the permutation index method is better than the baseline, but both methods do not scale well because of the increased contention and the overhead of final combination when all threads are used. On the Phi, they are the slowest methods when all threads are used, despite using much more memory. For the shared-buffer implementations, the permutation index method performs worse with multithreading because it leads to more severe contention over popular cache lines for atomic operations. The hybrid method addresses the contention problem. As shown in the figure, the hybrid methods perform the best. Using 256K as the threshold results in improved latency comparing to the 8K threshold (these threshold values correspond to L1 and L2 cache sizes), but uses more memory. Comparing the independent-buffer baseline approach, this method is 4.7x faster on the Skylake and 10.4x faster on the Phi, with much smaller memory usage.

To estimate the cost of building a permutation index, one can add the cost of an aggregation query ({\tt Q3}) and a materialization query ({\tt Q1}) using the baseline (rand) organization. In every configuration described above, building the index using all available threads would take just a few seconds.

\subsection{Results on Real Datasets}
\label{sec:exp:real}

We tested the permutation index method on three types of real-world datasets with different sizes and degrees of skew:
\begin{itemize}
  \item {\em Pageview}\footnote{\tt https://dumps.wikimedia.org/other/analytics} is a dataset of web request logs of wikipedia pages. The data is cleaned to filter out requests from search engine spiders, leaving only human traffic. In a database system, the request logs are stored as a fact table referencing a dimension table describing page information. For our experiment, we used the monthly request data from December 2018.
  \item {\em Product}\footnote{\tt https://s3.amazonaws.com/amazon-reviews-pds} is the Amazon reviews data. We believe the number of public product reviews is a proxy of the sales data, and it also exhibits skew. We consider the products as dimension data, and each review is a data item in the fact table. The data is separated into different categories, and each category exhibits different skewed data distribution. Our experiments uses eight categories with the most products.
  \item {\em Graph}\footnote{\tt https://snap.stanford.edu/data}. We tested several large graphs from the Stanford Large Network Dataset Collection~\cite{snapnets}, including social, communication, citation, and road networks. In our experiments, edges are stored as a fact table, with nodes stored in a dimension table. As an example, in a social network, nodes represent users and the dimension table describes user information. An aggregation query on the fact table, for instance, is to count the number of friends for every user (i.e., node degrees).
\end{itemize}
Table~\ref{tab:dataset} summarizes the sizes of the all the real-world datasets used in our experiments.

\begin{table}
  \centering
  \caption{Table Cardinality in Real-World Datasets}
  \label{tab:dataset}
  \begin{tabular}{|l|l|r|r|}                                      \hline
  Dataset         &             & Dimension   & Fact          \\  \hline\hline
  {\em Pageview}  & wikipedia   & 11,880,596  & 3,351,629,753 \\  \hline\hline
  {\em Product}   & book        & 2,717,050   & 19,531,329    \\  \hline
                  & dvd         & 221,086     & 5,069,140     \\  \hline
                  & ebook       & 1,292,480   & 17,622,415    \\  \hline
                  & home        & 918,287     & 6,221,559     \\  \hline
                  & music       & 675,893     & 4,751,577     \\  \hline
                  & pc          & 382,331     & 6,908,554     \\  \hline
                  & sports      & 753,280     & 4,850,360     \\  \hline
                  & wireless    & 767,830     & 9,002,021     \\  \hline\hline
  {\em Graph}     & friendster  & 65,608,366  & 1,806,067,135 \\  \hline
                  & orkut       & 3,072,441   & 117,185,083   \\  \hline
                  & livejournal & 4,847,571   & 68,993,773    \\  \hline
                  & pokec       & 1,632,803   & 30,622,564    \\  \hline
                  & youtube     & 1,134,890   & 2,987,624     \\  \hline
                  & wiki-talk   & 2,394,385   & 5,021,410     \\  \hline
                  & patent      & 3,774,768   & 16,518,948    \\  \hline
                  & ca-road     & 1,965,206   & 5,533,214     \\  \hline
  \end{tabular}
\end{table}

In the following experiments, we report the results of SIMD implementations for materialization and selection, and scalar implementations for aggregation (and heavy hitter aggregation). These implementations are generally fast as revealed in our microbenchmark analysis. For heavy hitter aggregation, we again compute the counts for the top 4,000 most frequent data items.

\begin{figure*}%[h]
  \begin{subfigure}{0.45\textwidth}
    \centering
    \includegraphics[width=\textwidth]{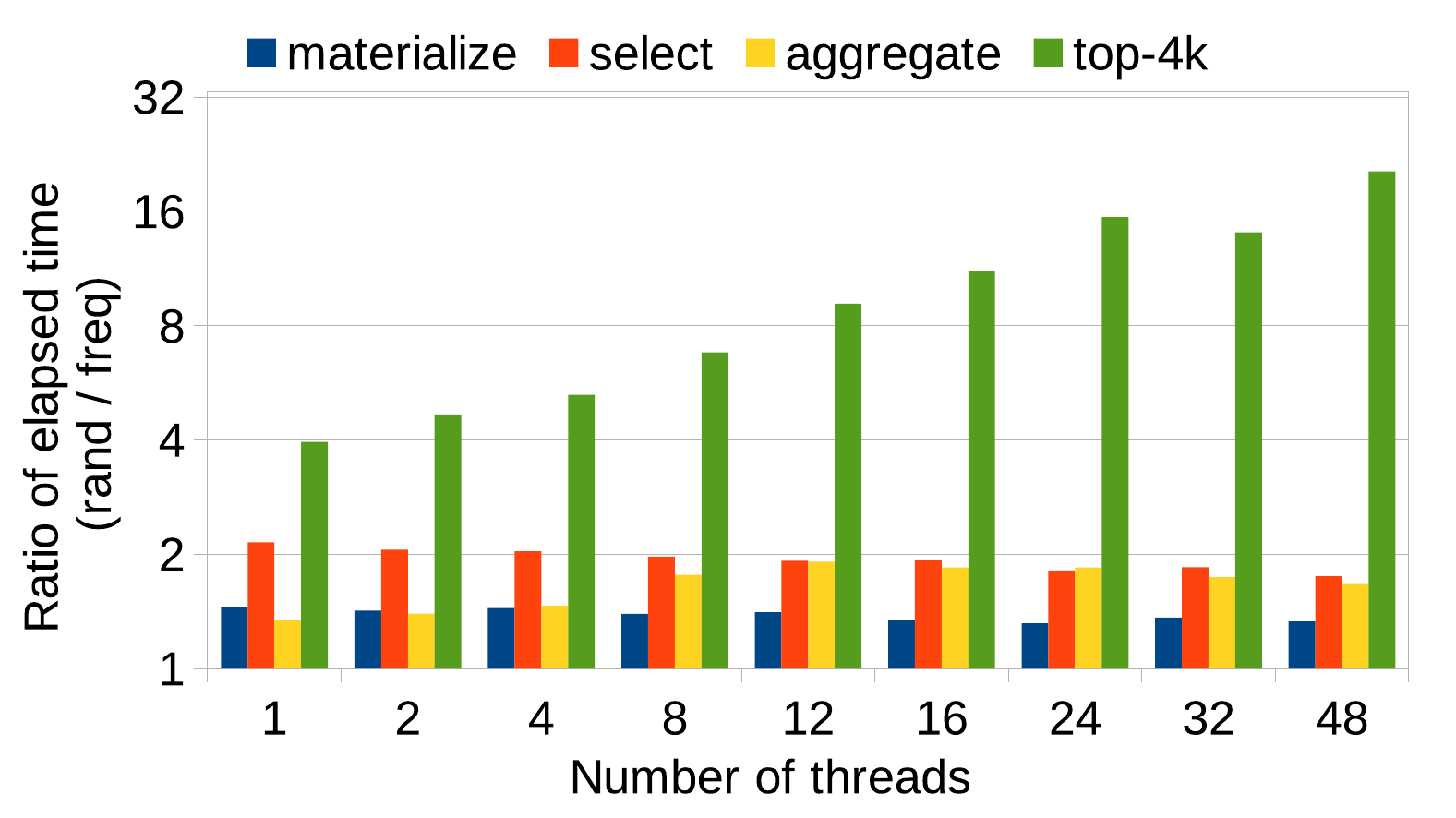}
    \caption{Skylake}
    \label{fig:exp:wiki:skylake}
  \end{subfigure}
  \hfill
  \begin{subfigure}{0.45\textwidth}
    \centering
    \includegraphics[width=\textwidth]{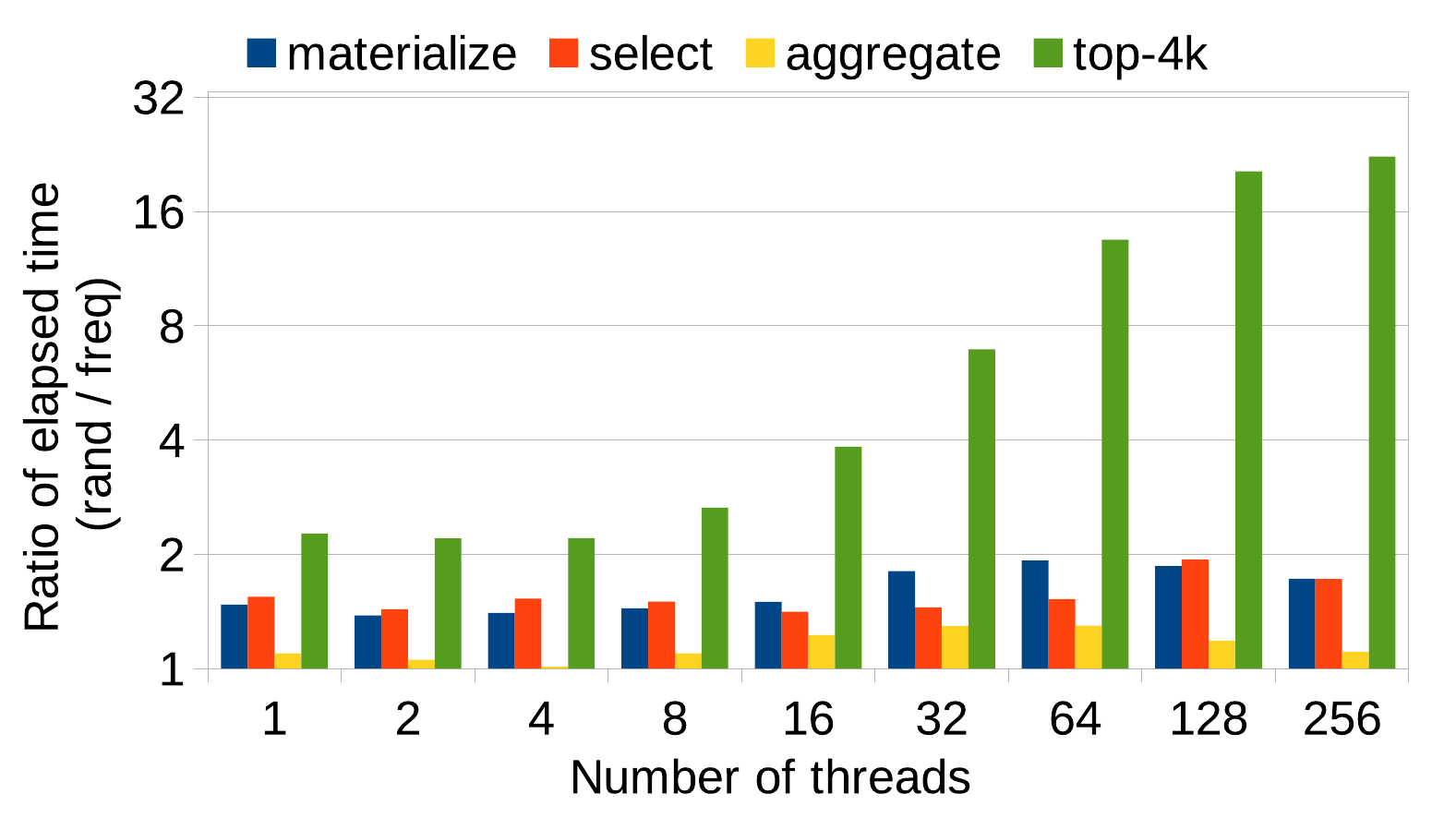}
    \caption{Phi}
    \label{fig:exp:wiki:phi}
  \end{subfigure}
  \caption{Performance speedup on {\em Pageview} using permutation indexes.}
  \label{fig:exp:wiki}
\end{figure*}

% \begin{figure}
%   \centering
%   \includegraphics[width=0.33\textwidth]{figures/wiki_skylake}
%   \caption{Performance speedup on {\em Pageview} (Skylake).}
%   \label{fig:exp:wiki:skylake}
% \end{figure}

Figure~\ref{fig:exp:wiki} shows the performance improvements on the largest dataset {\em Pageview}, using a varying number of threads. On the Skylake with a single thread, the permutation index speedup of materialization, selection, full aggregation, and heavy hitter aggregation is 1.5x, 2.2x, 1.3x, and 4.0x, respectively. The results on the Phi are generally slower than on the Skylake, but using permutation indexes similarly achieves 1.10--2.27x performance using a single thread. With more threads, permutation indexes speed up queries similarly for materialization, selection, and full aggregation, up to 2x. When all threads are used, the performance ratio for heavy hitter aggregation is 20.4x on the Skylake and 22.3x on the Phi. Because all threads can use their cache-resident private buffer to execute heavy hitter aggregation in parallel, the permutation index method is particularly effective.

Tables~\ref{tab:dataset:skylake} and~\ref{tab:dataset:phi} present the single-threaded latency results (in milliseconds) and performance speedups of using permutation indexes on all other datasets. On the Skylake for the product review and graph data, the performance improvements are up to 6.0x for heavy hitter aggregations, and up to 1.4--1.5x for other operations. Results on the Phi are generally slower, but they achieve similar speedups using permutation indexes.

\begin{table*}%[h]
  \centering
  \caption{Performance Results with Real-World Datasets on Skylake}
  \label{tab:dataset:skylake}
  \begin{tabular}{|l|r|r|r|r|r|r|r|r|r|r|r|}
  \hline
    & Materialize & \  & \  & Select & \  & \  & Aggregate & \  & \  & Top-4k & \  \\ \hline
    \hline
    % wikipedia & 14608.2 & 10051.7 & 1.45 & 5691.21 & 2644.98 & 2.15 & 15713.0 & 11707.7 & 1.34 & 3972.64 & 3.96 \\ \hline
    % \hline
    {\em Product} & rand & freq & ratio & rand & freq & ratio & rand & freq & ratio & freq & ratio \\ \hline
    book & 74.89 & 53.54 & 1.40 & 11.88 & 10.07 & 1.18 & 71.40 & 62.23 & 1.15 & 25.46 & 2.80 \\ \hline
    dvd & 5.57 & 4.61 & 1.21 & 1.77 & 1.75 & 1.01 & 5.47 & 4.84 & 1.13 & 5.34 & 1.02 \\ \hline
    ebook & 49.04 & 35.53 & 1.38 & 10.10 & 8.20 & 1.23 & 53.09 & 39.48 & 1.34 & 20.76 & 2.56 \\ \hline
    home & 14.40 & 10.46 & 1.38 & 3.10 & 2.53 & 1.23 & 14.38 & 10.60 & 1.36 & 6.99 & 2.06 \\ \hline
    music & 10.23 & 7.49 & 1.37 & 2.18 & 1.88 & 1.16 & 10.24 & 7.70 & 1.33 & 5.61 & 1.82 \\ \hline
    pc & 10.96 & 7.60 & 1.44 & 2.95 & 2.80 & 1.05 & 10.19 & 6.82 & 1.49 & 6.99 & 1.46 \\ \hline
    sports & 10.13 & 7.64 & 1.33 & 2.41 & 1.95 & 1.23 & 10.04 & 6.87 & 1.46 & 5.47 & 1.84 \\ \hline
    wireless & 19.96 & 13.46 & 1.48 & 4.62 & 3.67 & 1.26 & 21.61 & 13.38 & 1.62 & 9.78 & 2.21 \\ \hline
    \hline
    {\em Graph} & rand & freq & ratio & rand & freq & ratio & rand & freq & ratio & freq & ratio \\ \hline
    friendster & 16514.6 & 12136.3 & 1.36 & 5243.83 & 3753.34 & 1.40 & 16423.1 & 12692.4 & 1.29 & 2718.59 & 6.04 \\ \hline
    orkut & 400.85 & 386.93 & 1.04 & 73.30 & 69.67 & 1.05 & 459.44 & 440.40 & 1.04 & 169.06 & 2.72 \\ \hline
    livejournal & 248.14 & 230.95 & 1.07 & 47.31 & 44.02 & 1.07 & 276.49 & 267.41 & 1.03 & 99.58 & 2.78 \\ \hline
    pokec & 100.37 & 88.78 & 1.13 & 18.34 & 16.66 & 1.10 & 112.33 & 101.78 & 1.10 & 44.56 & 2.52 \\ \hline
    youtube & 6.23 & 4.32 & 1.44 & 1.45 & 1.12 & 1.30 & 6.00 & 4.53 & 1.32 & 3.46 & 1.73 \\ \hline
    wiki-talk & 11.25 & 9.46 & 1.19 & 2.54 & 1.95 & 1.30 & 10.07 & 9.38 & 1.07 & 5.00 & 2.02 \\ \hline
    patent & 59.39 & 55.11 & 1.08 & 10.10 & 9.92 & 1.02 & 62.32 & 60.71 & 1.03 & 24.44 & 2.55 \\ \hline
    ca-road & 16.94 & 16.57 & 1.02 & 2.95 & 2.9 & 1.02 & 16.46 & 16.19 & 1.02 & 8.37 & 1.97 \\ \hline
  \end{tabular}
\end{table*}

\begin{table*}
  \centering
  \caption{Performance Results with Real-World Datasets on Phi}
  % \todo{update amazon results}
  \label{tab:dataset:phi}
  \begin{tabular}{|l|r|r|r|r|r|r|r|r|r|r|r|}
  \hline
    & Materialize & \  & \  & Select & \  & \  & Aggregate & \  & \  & Top-4k & \  \\ \hline
    \hline
    {\em Product} & rand & freq & ratio & rand & freq & ratio & rand & freq & ratio & freq & ratio \\ \hline
    book & 218.36 & 167.07 & 1.31 & 49.28 & 49.11 & 1 & 289.77 & 277.2 & 1.05 & 141.81 & 2.04 \\ \hline
    dvd & 21.45 & 19.39 & 1.11 & 8.76 & 8.58 & 1.02 & 36.98 & 33.73 & 1.10 & 36.58 & 1.01 \\ \hline
    ebook & 145.41 & 105.27 & 1.38 & 43.62 & 39.08 & 1.12 & 248.39 & 220.59 & 1.13 & 125.68 & 1.98 \\ \hline
    home & 50.97 & 35.06 & 1.45 & 15.85 & 13.67 & 1.16 & 87.07 & 78.32 & 1.11 & 44.62 & 1.95 \\ \hline
    music & 36.59 & 26.83 & 1.36 & 9.98 & 9.12 & 1.09 & 64.68 & 58.57 & 1.10 & 34.60 & 1.87 \\ \hline
    pc & 36.06 & 29.73 & 1.21 & 15.64 & 13.81 & 1.13 & 75.33 & 57.59 & 1.31 & 49.75 & 1.51 \\ \hline
    sports & 39.08 & 26.9 & 1.45 & 12.13 & 10.48 & 1.16 & 68.68 & 59.3 & 1.16 & 35.29 & 1.95 \\ \hline
    wireless & 60.45 & 45.23 & 1.34 & 21.69 & 18.81 & 1.15 & 118.96 & 98.45 & 1.21 & 66.18 & 1.80 \\ \hline
    \hline
    {\em Graph} & rand & freq & ratio & rand & freq & ratio & rand & freq & ratio & freq & ratio \\ \hline
    friendster & 27300.2 & 26776.80 & 1.02 & 25146.1 & 17061.5 & 1.47 & 30691.60 & 28938.5 & 1.06 & 12902.7 & 2.38 \\ \hline
    orkut & 1627.33 & 1483.74 & 1.10 & 303.01 & 290.24 & 1.04 & 1890.73 & 1763.08 & 1.07 & 842.5 & 2.24 \\ \hline
    livejournal & 978.51 & 855.56 & 1.14 & 231.75 & 178.46 & 1.3 & 1140.16 & 1081.29 & 1.05 & 495.52 & 2.30 \\ \hline
    pokec & 404.86 & 316.48 & 1.28 & 77.64 & 77.41 & 1 & 488.34 & 465.34 & 1.05 & 220.46 & 2.22 \\ \hline
    youtube & 30.29 & 21.08 & 1.44 & 7.72 & 7.29 & 1.06 & 45.7 & 41.68 & 1.10 & 21.88 & 2.09 \\ \hline
    wiki-talk & 44.36 & 33.09 & 1.34 & 13.13 & 12.2 & 1.08 & 71.82 & 68.61 & 1.05 & 36.36 & 1.97 \\ \hline
    patent & 234.93 & 223.26 & 1.05 & 44.99 & 41.64 & 1.08 & 266.96 & 261.31 & 1.02 & 117.75 & 2.27 \\ \hline
    ca-road & 76.14 & 75.23 & 1.01 & 12.76 & 12.76 & 1 & 89.91 & 85.75 & 1.05 & 41.31 & 2.18 \\ \hline
  \end{tabular}
\end{table*}

\section{Related Work}
\label{sec:rel}

Skew has been used to improve OLAP query execution in several ways that are orthogonal to those described in this paper. High-frequency data items can be represented with fewer bits to reduce data transfer costs~\cite{li2015padded,hentschel2018column}. When skew causes load imbalances in a parallel join, explicit scheduling of common keys can overcome this imbalance~\cite{wolf1991effective, dewitt1992practical,duggan2015skew}. At lower levels of the memory hierarchy, popular rows may be surrounded by cold rows in a disk page. This observation has led some systems to adopt row caches that keep popular rows in RAM even if their page has been evicted from the buffer pool~\cite{lastovica2003guide}. Such a solution is not viable at higher levels of the memory hierarchy because the programmer has little direct control of what data is loaded into the processor caches.

Techniques for tracing data access streams to quantify the degree of latent locality have been described~\cite{chilimbi2001efficient}. Distributional parameters of skewed data streams can be estimated~\cite{korn2006modeling}. Skew can affect the performance of selectivity estimators in query optimization, and skew-aware methods are required for accurate estimation~\cite{lynch1988selectivity}.

The advent of large main memory capacity paved the way for high-performance OLAP query execution. Column-oriented execution \cite{manegold2000optimizing} and cache-conscious operators \cite{manegold2002optimizing} were proposed before the advent of multi-core CPUs. Analytical database systems adopted column-oriented storage, while focusing on compression \cite{raman2013db2, stonebraker2005c, willhalm2009simd} and complex materialization strategies \cite{abadi2007materialization, shrinivas2013materialization} to further optimize memory access. Block-at-a-time execution \cite{boncz2005monetdb} and code generation \cite{gupta2015amazon,kim2010fast,neumann2011efficiently, tahboub2018architect} are both state-of-the-art designs for analytical query engines~\cite{kersten2018everything}.

SIMD optimizations have been applied to isolated operators using key--rid pairs. Many join implementations exist \cite{balkesen2013multi, balkesen2013main, blanas2011design, kim2009sort, schuh2016experimental}, including for many-core platforms \cite{jha2015improving}. SIMD implementations of stand-alone operators such as sorting \cite{chhugani2008efficient, inoue2007aa, polychroniou2014comprehensive, satish2010fast} are also common. Linear-access operators such as scans \cite{zhou2002implementing} and compression \cite{lang2016data, polychroniou2015efficient, willhalm2009simd}, are inherently data-parallel. Advanced SIMD optimizations \cite{polychroniou2014vectorized, polychroniou2015rethinking} include non-linear-access operators. A recently proposed vector algebra \cite{pirk2016voodoo} proposes templates for both SIMD CPUs and SIMT GPUs. Non-linear-access operators are wrappers to vector gathers, such as joins executed by dereferencing join indexes, which are likely to benefit from the permutation indexes when there is skew in the join attribute distribution.

% In contrast to OLAP, on-line transaction processing (OLTP) workloads stress transactional update peformance, and their queries touch a relatively small portion of the database. In-memory processing is also used to speed up such workloads~\cite{kallman2008h, debrabant2013anti, larson2013hekaton, zhang2016reducing}. The different performance goals can lead to very different system architectures~\cite{larson2013hekaton}, although some memory-optimized systems support both OLAP and OLTP workloads~\cite{sikka2012efficient, kemper2011hyper, pavlo2017self}.

Other architecture-specific optimizations such as software prefetching and query compilation are used to improve database performance \cite{menon2017relaxed}. Cache partitioning is used to address the cache pollution problem in shared cache to accelerate concurrent workloads \cite{noll2018accelerating}. The performance of frequent pattern mining algorithms can also be improved by tuning data layout and access patterns \cite{wei2007programming}.
\section{Conclusions}

We propose permutation indexes to reorder data so that popular data items are concentrated in the cache hierarchy. Using this method, we can exploit the data skew inherent in many practical domains to improve cache utilization at all levels. Efficient database operators can be developed using permutation indexes, combining the benefits of cache optimization with SIMD vectorization. Our method is also effective when working together with multithreaded execution and software prefetching, further speeding up query execution. Through extensive experiments with real and synthetic data, we demonstrate that the performance of materialization, selection, and aggregation queries can be significantly improved.

% \section*{Acknowledgment}

\bibliographystyle{IEEEtran}
\bibliography{IEEEabrv, paper}

\end{document}